\documentclass[twoside,11pt]{article}

\usepackage{blindtext}

\usepackage[abbrvbib,preprint]{jmlr2e}
\usepackage{amsmath}
\usepackage{multirow}
\usepackage{mathtools}

\DeclareMathOperator*{\argmin}{argmin}

\usepackage{lastpage}

\ShortHeadings{Functional Conformal Anomaly Detection}{Adams, Berman, Michalenko, and Tucker}
\firstpageno{1}

\begin{document}

\title{Conformal Anomaly Detection for Functional Data with Elastic Distance Metrics}

\author{\name Jason Adams \email jradams@sandia.gov \\
       \addr Department of Statistical Sciences\\
       Sandia National Laboratories\\
       Albuquerque, NM 87123, USA
       \AND
       \name Brandon Berman \email bjberma@sandia.gov \\
       \addr Department of Statistical Sciences\\
       Sandia National Laboratories\\
       Albuquerque, NM 87123, USA \\
       \name Joshua Michalenko \email jjmich@sandia.gov \\
       \addr Department of Proliferation Signature and Data Exploitation\\
       Sandia National Laboratories\\
       Albuquerque, NM 87123, USA \\
       \name J. Derek Tucker \email jdtuck@sandia.gov \\
       \addr Department of Statistical Sciences\\
       Sandia National Laboratories\\
       Albuquerque, NM 87123, USA}

\maketitle

\begin{abstract}  
This paper considers the problem of outlier detection in functional data analysis focusing particularly on the more difficult case of shape outliers. We present an inductive conformal anomaly detection method based on elastic functional distance metrics. This method is evaluated and compared to similar conformal anomaly detection methods for functional data using simulation experiments. The method is also used in the analysis of two real exemplar data sets that show its utility in practical applications. The results demonstrate the efficacy of the proposed method for detecting both magnitude and shape outliers in two distinct outlier detection scenarios.

\end{abstract}


\section{Introduction}
\label{intro}

Functional data are prevalent in many applications across a wide range of scientific domains \citep{ullah2013}. As such, methods enabling the principled statistical analysis of functional data are important. The development and application of such methods have been rich areas of research for many years, and functional data analysis (FDA) is now a well-established branch of statistics \citep{Ramsay2005, srivastava2016, kokoszka2017}. This paper introduces a novel approach to detecting anomalous functional data. Our method leverages both the conformal prediction (CP) framework \citep{vovk2005} and the elastic functional data analysis (EFDA) framework \citep{srivastava2016}. Technical details necessary for understanding these frameworks and why they are useful for anomaly detection are provided in Section \ref{prelims}.

Functional data vary continuously over some independent variable or variables. In this work, we only consider univariate functional data where all functions are observed over a single independent variable. Within a given set of $n$ observed functions, we assume that each function is observed over the same values of the independent variable. While this assumption is often not satisfied in practice, smoothing and interpolation methods can be used to adjust the observed data for this purpose. Although this is far from a trivial process, such methods for preprocessing functional data are outside the scope of this paper. We refer readers to \cite{Ramsay2005} as a starting point for smoothing and interpolation methods.

Throughout this paper, the terms \emph{anomaly} and \emph{outlier} are used interchangeably to describe an observation that does not fit well within a distribution. This is in line with the definition used by \cite{aggarwal2016} which says that ``[a]n outlier is a data point that is significantly different from the remaining data." Within this definition, there are two distinct situations we consider. In the first scenario, a representative, outlier-free data set is available and potentially suspect observations from an external source are compared to the pristine data. In the second scenario, we seek to identify the most outlying observations within a single data set. While the method we introduce is most appropriately used in the first scenario, both scenarios are considered in the simulation experiments and analysis of exemplar data sets in Sections \ref{sim} and \ref{exem}.

\subsection{Related Work} \label{intro:litrev}
Much previous work has been done in both functional outlier detection and conformal anomaly detection. This section reviews some of the most relevant methods and compares them to the present work.

\subsubsection{Functional Outlier Detection} \label{intro:litrev:fnout}

Numerous methods for functional outlier detection have been proposed in the literature.  Many papers distinguish between \emph{magnitude} and \emph{shape} outliers. As described by \cite{hyndman2010}, magnitude outliers ``lie outside the range of the vast majority of the data" while shape outliers ``may be within the range of the rest of the data but have a very different shape from other curves." Visualization methods are often sufficient to identify magnitude outliers, but shape outliers are typically much more difficult to identify.

Several approaches to functional outlier detection depend on visualization, aiming to extend standard univariate and multivariate techniques, such as boxplots or bagplots \citep{rousseeuw1999}, to the functional case. The primary challenge with such an extension is to determine a reasonable method for ordering functional data. A robust principal component (PC) analysis is used by \cite{hyndman2010} to reduce the dimensionality of the functional data. Using the first two PCs, depth and density measures are used to order the PC scores. From these orderings, three functional visualization tools — the rainbow plot, the bagplot, and the boxplot - are constructed and used to detect both magnitude and shape outliers. Similarly, \cite{sun2011} use the concept of band depth from \cite{lopez2009} to construct functional boxplots, and \cite{huang2019} introduce the notion of total variation depth to order functional data and construct visualizations for detecting both magnitude and shape outliers. \cite{arribas2014} propose visualization tools called outliergrams, which are constructed based on metrics from \cite{lopez2011}. Noting that magnitude outliers are much easier to detect than shape outliers, \cite{dai2020} apply several transformations to functional data so that shape outliers can be detected as magnitude outliers. 

Other methods of functional outlier detection do not rely on visualization. \cite{sawant2012} develop a robust principal component method and apply a multivariate outlier detection method from \cite{hubert2005} to the PC scores. Similarly, \cite{yu2017} represent functional data with a B-spline basis \citep{Ramsay2005} and use the minimum covariance determinant method \citep{rousseeuw1999mcd} on the basis coefficients to detect outliers. Both of these methods can be viewed as using standard outlier detection methods on a reduced number of features extracted algorithmically from a set of functional data. In a different approach by \cite{azcorra2018}, three features are selected to measure the degree of outlyingness in terms of magnitude, shape, and amplitude respectively (note that this paper further divides what other papers call shape outliers into both shape and amplitude outliers). Thresholds are then set on these features to identify outliers.

Additionally, several methods for functional outlier detection within the EFDA framework have also been proposed. In the method proposed by \cite{xie2017}, elastic distance metrics (discussed in more detail in Section \ref{prelims} in the present work) are used to construct functional boxplots. Measures of elastic depths are introduced by \cite{harris2021} and used to identify shape outliers. While not explicitly presented as an outlier detection method, \cite{tucker2020} introduced functional tolerance bounds that are constructed by bootstrapping and using the EFDA boxplots of \cite{xie2017}, and these tolerance bounds can be used to identify outliers.

While the present work does make use of the EFDA framework and can be used for detecting both magnitude and shape outliers, it differs from the above-mentioned methods in its reliance on the conformal prediction framework, which none of the others use. It also does not rely on either visualization or dimension reduction/featuriziation techniques. 

\subsubsection{Conformal Anomaly Detection} \label{intro:litrev:cad}
Within the conformal prediction literature, a number of papers have focused on the problem of detecting outliers. The method of conformal anomaly detection (CAD) was introduced by \cite{laxhammar2014}. In a follow up work \citep{laxhammar2015}, the method of inductive conformal anomaly detection (ICAD) was proposed, and both of these focused on trajectory data. Since we will frame our proposed method as ICAD for functional data, more details on ICAD are given in Section \ref{prelims}. In the work of \cite{cai2022}, ICAD is used to detect outliers in cyber-physical systems, such as autonomous vehicles, for the purpose of improving the control mechanisms of such systems. An adaption of ICAD for univariate time series is proposed by \cite{ishimtsev2017}. A method for adjusting the outputs of ICAD algorithms to reduce the number of false positives (i.e., incorrectly labeling an observation as an outlier) is presented by \cite{bates2023}. The clearest distinction between these papers and the present work is that none of them are concerned with functional data.

Some other works deal with outlier detection using the CP framework that are somewhat less related to the present paper. First, \cite{liang2022} use labeled outliers to adaptively weight conformal p-values, which are then used to determine whether to label a test observation as an outlier or inlier. We do not assume to have labeled outliers in the present work. The focus of \cite{guan2022} is on CP for classification problems. The authors also introduce a method for estimating the expected proportion of outliers that are correctly labeled as such. We instead focus solely on the problem of outlier detection and do not consider classification problems herein. Lastly, \cite{haroush2021} proposed a method for detecting outliers with deep neural networks. The authors only mention in passing that their approach can be viewed within the CP framework and present their method instead as statistical hypothesis testing. 

\subsubsection{Functional Conformal Anomaly Detection} \label{intro:litrev:fncad}
To date, there have only been a small number of papers that use conformal prediction within a functional data context. In the work of \cite{wang2025}, EFDA methods are used with CP to provide bounds for partially observed functional data. Both \cite{diana2023} and \cite{de2024} use CP for spatial functional data. CP has been applied to multivariate functional data and functional time series data by \cite{diquigiovanni2022} and \cite{ajroldi2023}, respectively. However, two previous works focus on the functional outlier detection problem. First, \cite{lei2015} introduce a CP approach that uses a Gaussian mixture density fit on a basis representation of functional data to visualize classes of functional data. While the main focus of the paper is on visualization and data exploration, the method is also useful for outlier detection. In the work of \cite{diquigiovanni2021}, a CP method is proposed for detecting outliers in functional data that is based on distance to a mean function and local variability of the functional data. 

Given the close relationship between the methods presented by \cite{lei2015} and \cite{diquigiovanni2021} and the one we are proposing, we give a detailed explanation of them in Section \ref {efdm} and employ them as benchmark methods in Section \ref{sim}. We note that these methods were not originally presented as ICAD methods, but it is natural to frame them as such. Throughout, we will refer to the ICAD version of the method from \cite{lei2015} as GMD (for Gaussian Mixture Density) and the ICAD version of the method from \cite{diquigiovanni2021} as SNCM (for Supremum Non-conformity Measure). The primary difference between our proposed method and GMD and SNCM is that we base our method on the EFDA framework.

\subsection{Contribution and Outline} \label{intro:outline}
Our primary contribution in this work is to introduce a novel ICAD method, based on the EFDA framework, for detecting functional data outliers. This method most naturally fits the first outlier detection scenario, where a pristine data set is available, and we seek to detect outliers in new or external data. Furthermore, we propose a leave-one-out approach to address the second outlier detection scenario, which involves identifying the most anomalous observations within a single data set. We conduct empirical evaluations of these approaches using simulated data and compare our method to both GMD and SNCM. Our results demonstrate the efficacy of our method in both outlier detection scenarios.

The outline of the paper is as follows: Section \ref{prelims} provides necessary details regarding EFDA, CP, and ICAD to understand our method. In Section \ref{efdm}, we introduce our proposed method. Section \ref{sim} describes our empirical method comparison. Our method is demonstrated on two real data exemplars in Section \ref{exem}, and we conclude in Section \ref{conclusion}.

\section{Preliminaries} \label{prelims}
This section gives necessary details to understand our proposed method.

\subsection{Elastic Functional Data Analysis} \label{prelims:efda}
\begin{figure}
    \centering
    \begin{tabular}{cc}
    \includegraphics[width=0.5\linewidth]{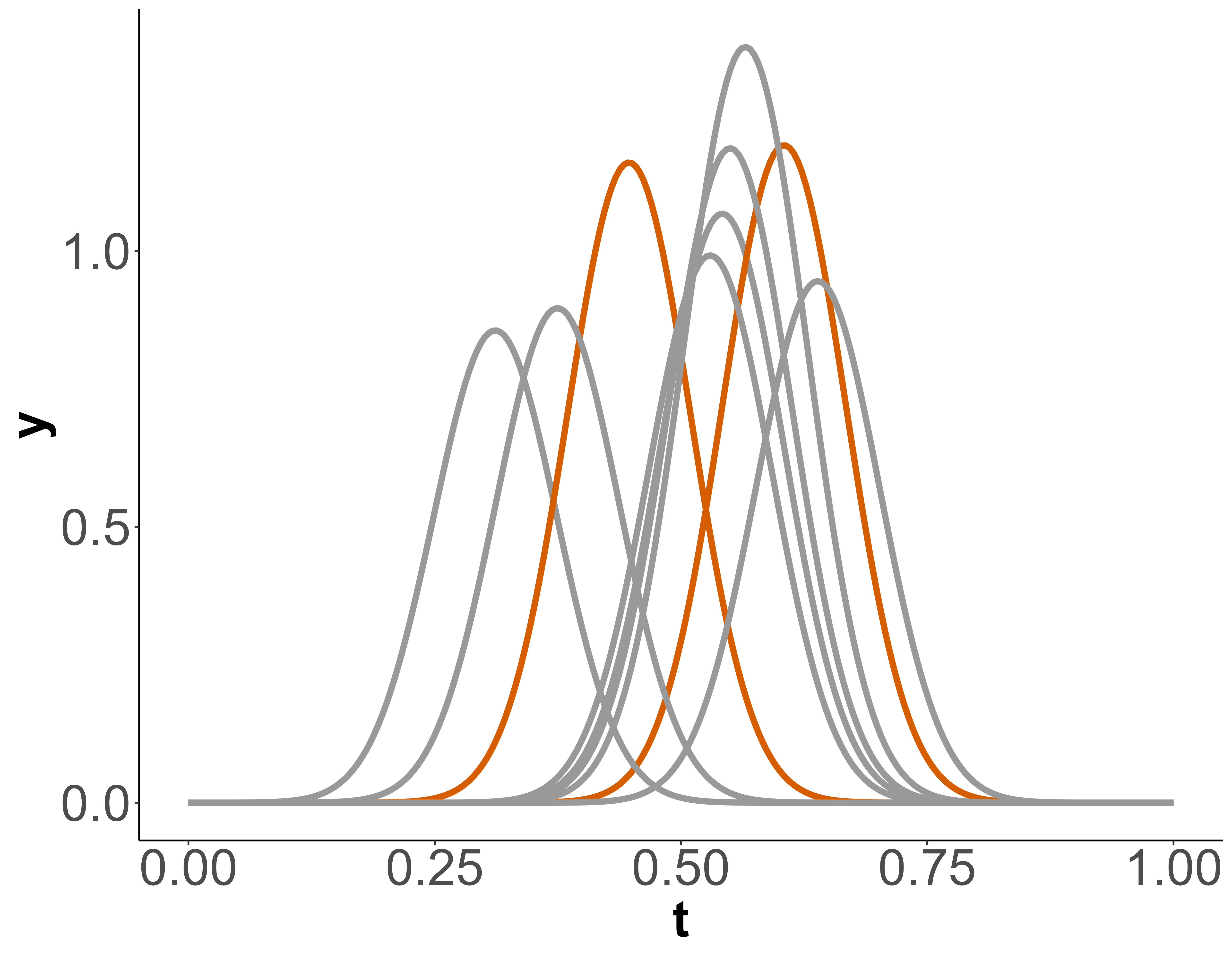}   &  \includegraphics[width = 0.5\linewidth]{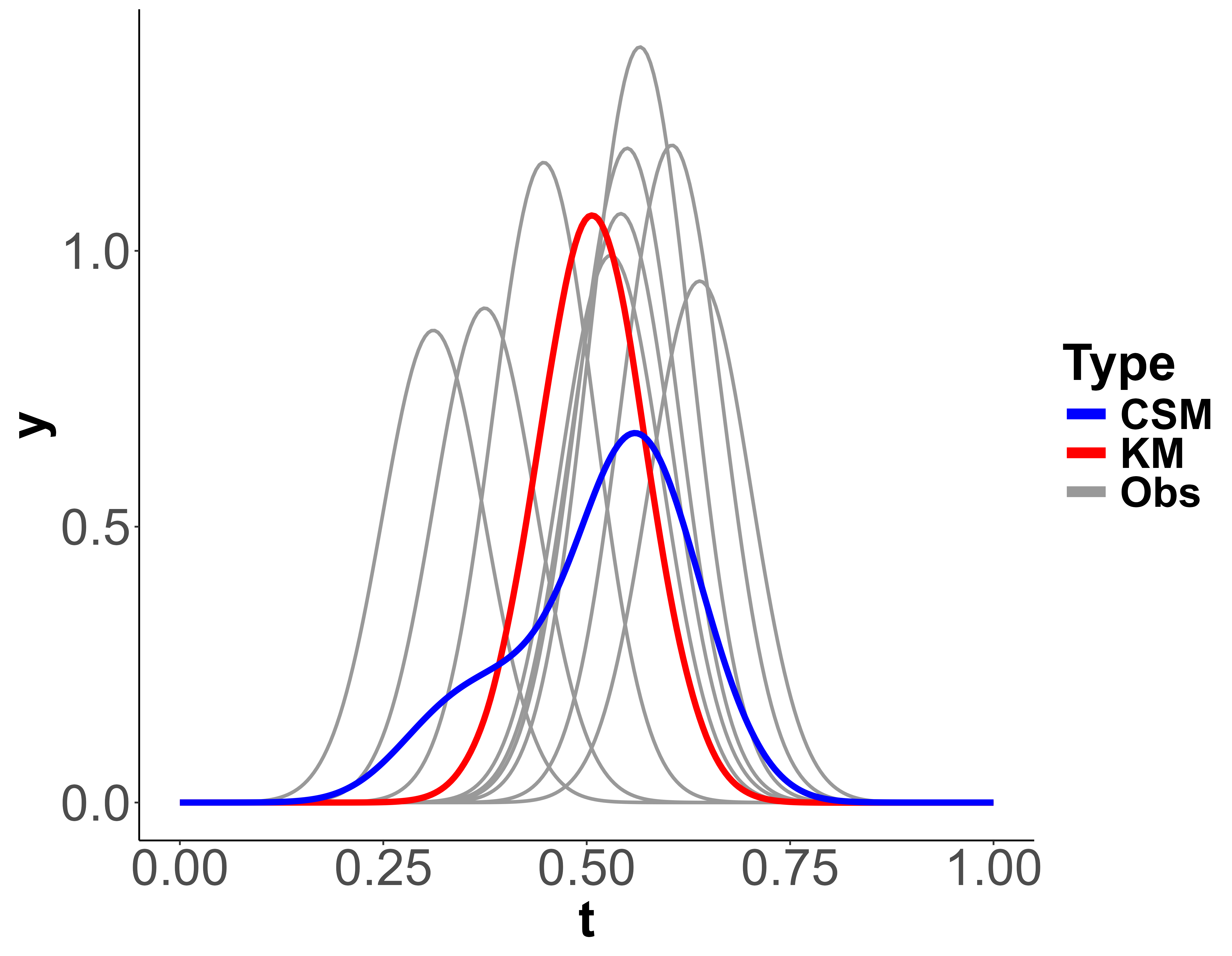} \\
    (a) & (b)
    \end{tabular}
    \caption{(a) A small sample of example functional data. (b) The same functional data observations (OBS) in grey with the Karcher mean (KM) and cross-sectional mean (CSM) overlaid in red and blue, respectively.}
    \label{fig:Sec2_exfndat}
\end{figure}

In the analysis of functional data, there are two aspects of variability that must be considered. These are \emph{phase} (or $x$-axis) variability and \emph{amplitude} (or $y$-axis) variability. Figure \ref{fig:Sec2_exfndat}(a) shows a sample of nine functional data observations that display both phase and amplitude variability. For instance, the two highlighted functions in Figure \ref{fig:Sec2_exfndat}(a) vary considerably in phase but only very little in amplitude. The EFDA framework provides distance metrics that quantify the degree of separation between two functions with regard to both amplitude and phase. It also allows for estimating a Karcher mean, which is a better measure of center than a standard cross-sectional mean \citep{Ramsay2005} when functional data contain phase variability. In Figure \ref{fig:Sec2_exfndat}(b), the grey curves are the same functional data as in panel (a), the red function is the Karcher mean, and the blue function is the cross-sectional mean. Clearly, the Karcher mean provides a much better representation of the average function than the cross-sectional mean as it better represents the peakedness, the symmetry, and spread of functions present in the data.

Because the math underpinning the EFDA framework is quite involved, we provide only the essential details for computing amplitude and phase distances as well as the Karcher mean for a set of functional data. We refer readers who are interested in more detail on the EFDA framework to \cite{srivastava2016}.

Let $f$ be a real-valued and absolutely continuous function over the domain $[0,1]$ and $\mathcal{F}$ be the set of all such functions\footnote{In practice, the function $f$ can be defined over any closed interval of the form $[a,b] \subset \mathbb{R}$. This is typically done by simply rescaling the interval $[0,1]$ to $[a,b]$.}. Let $\gamma: [0,1] \rightarrow [0,1]$ be a boundary preserving diffeomorphism with $\gamma(0) = 0$ and $\gamma(1) = 1$. Further, let $\Gamma$ be the set of all such diffeomorphisms. We call $\gamma$ a \emph{warping function} as the composition of $f$ and $\gamma$, $f\circ\gamma$, effectively warps the function $f$ but maintains the original domain over which $f$ is defined. Also, we denote the \emph{square root slope function} (SRSF) of a function $f$ by $q:[0,1] \rightarrow \mathbb{R}$ such that $q(t) = \mathrm{sign}(\dot{f}(t))\sqrt{|\dot{f}(t)|}$ where $\dot{f}$ is the derivative of $f$. Similarly, we denote the SRSF of the warping function to be $\psi$. However, since $\gamma>0$ and $\dot{\gamma}>0$ then the SRSF of $\psi$ is simplified to $\psi=\sqrt{\dot{\gamma}}$.

\subsubsection{Elastic Distance Metrics} \label{prelims:efda:dists}
Let $f_1$, $f_2 \in \mathcal{F}$ and $q_1$, $q_2$ be their corresponding SRSFs. Then the amplitude distance between $f_1$ and $f_2$ is defined as 
\begin{align}
    d_a(f_1, f_2) = \inf_{\gamma \in \Gamma} ||q_1 - (q_2 \circ \gamma)\sqrt{\dot{\gamma}}||
    \label{eq:2_1}
\end{align}
where $||\cdot||$ represents the standard $\mathbb{L}^2$ metric, which is the Fisher-Rao metric and a proper distance. If we compute this distance using $f$ directly, then it is not a proper distance and hence the reason that the SRSF transformation is used. The reader is referred to \cite{srivastava2016} for further details. The warping function, $\gamma$, which optimizes the distance in Equation \ref{eq:2_1} is typically identified in practice through the Dynamic Programming algorithm \citep{bertsekas2012}. Intuitively, the optimal warping function, $\gamma$, not only minimizes the distance from $q_2$ to $q_1$, but also effectively aligns $f_2$ to $f_1$ as $(q_2\circ\gamma)\sqrt{\dot{\gamma}}$ is the SRSF of $f_2\circ\gamma$. For all EFDA computations herein, we use the \texttt{fdasrvf} package (version 2.3.4) \citep{tucker2017} within the \texttt{R} programming language \citep{R2023}.

For two warping functions, $\gamma_1$, $\gamma_2 \in \Gamma$, the phase distance between them is computed as 
\begin{align}
    d_\gamma(\gamma_1, \gamma_2) = \cos^{-1}\left(\int_0^1 \psi_1(t)\psi_2(t)dt\right)
\end{align}
where $\psi_1$ and $\psi_2$ are the SRSFs of $\gamma_1$ and $\gamma_2$, respectively. In order to find the phase distance between two functions, $f_1,f_2\in\mathcal{F}$, we first find the optimal warping function from $f_2$ to $f_1$. Again, denote this function as $\gamma$. Next the warping function from $f_1$ to itself is simply the identity function ($\gamma_{I}(t)=t$), and the SRSF of the identity function is the constant function equal to one ($\sqrt{\dot{\gamma_{I}}(t)}=1$). Thus, the phase distance between $f_2$ and $f_1$ simplifies to 
\begin{align}
    d_p(f_1, f_2) = d_\gamma(\gamma_I, \gamma) = \cos^{-1}\left(\int_0^1 \psi(t)dt\right)
\end{align}
where $\psi$ is the SRSF of the optimal warping function $\gamma$ that aligns $f_2$ to $f_1$. This simplification occurs because the space of all $\psi$s forms a Hilbert Sphere (see \citealp{tucker-wu-srivastava:2013}).

\subsubsection{Karcher Mean} \label{prelims:efda:km}
Let $f_1, \ldots, f_n$ represent a set of functions from the function space $\mathcal{F}$ and $q_1, \ldots, q_n$ be their respective SRSFs. The Karcher mean of $f_1, \ldots, f_n$ is given by
\begin{align}\label{eq:sec2_kmf}
    \mu_f = \argmin_{f \in \mathcal{F}}\sum_{i=1}^n d_a(f, f_i)^2.
\end{align}%
Equivalently, we can define the Karcher mean of the SRSFs, $q_1, \ldots, q_n$ as 
\begin{align}\label{eq:sec2_kmq}
    \mu_q = \argmin_{q \in \mathbb{L}^2} \sum_{i=1}^n \left(\inf_{\gamma_i \in \Gamma} ||q - (q_i \circ \gamma_i)||^2 \right).
\end{align}
Note that $\mu_q$ is the SRSF of $\mu_f$. The algorithm used in the \texttt{fdasrvf R} package finds $\mu_q$ and then transforms it to $\mu_f$. 

The transformation from SRSF space back to the original function space, assuming a generic function $f$ and its SRSF $q$, is \begin{equation}\label{eq:sec2_reconstructfromq}f(t) = f(t_0) + \int_{t_0}^t q(s)|q(s)|ds\end{equation} where $f(t_0)$ is the function value at the initial time point, $t_0$. When obtaining $\mu_f$ from $\mu_q$, the initial value is computed as $n^{-1}\sum_{i=1}^n f_i(t_0)$.

\subsection{Conformal Prediction and ICAD} \label{prelims:cp}
Conformal prediction was first introduced by \cite{vovk2005} as a distribution-free approach to obtaining valid uncertainty quantification (UQ). It has recently been gaining popularity in the machine learning literature as a computationally efficient means to obtain high quality UQ with many data types and different classes of models \citep{angelopoulos2021, zhou2024}. Let $(X_1,Y_1), \ldots, (X_n,Y_n) \sim \mathcal{P}_{\mathcal{X}\times\mathcal{Y}}$ be an \emph{i.i.d} random sample and $X_{n+1}$ be a test point and $Y_{n+1}$ the unobserved label associated with the test point $X_{n+1}$. For simplicity, we use the notation $Z_i = (X_i, Y_i)$. CP methods produce a \emph{prediction set}, $C(X_{n+1})$, for the test point $X_{n+1}$, such that $P(Y_{n+1} \in C(X_{n+1})) \ge 1-\alpha$ for any level of significance, $\alpha \in (0,1)$. This property is known as the marginal coverage guarantee, and we note that it also holds under the weaker assumption of exchangeability of $Z_1, \ldots, Z_n$.

To produce a prediction set, a \emph{non-conformity measure} (NCM) must first be defined. The NCM is a function that measures how well $Z_i$ conforms with the remaining observations. Common choices for the NCM are functions that are inherent measures of uncertainty like a residual from a regression model or a nearest-neighbors distance. The original formulation of CP was carried out in what is called the \emph{transductive} method. In this approach, NCM values are computed for all $Z_i$, $i\in\{1,\ldots,n,n+1\}$, relative to the set $\mathcal{Z}_{(-i)} = \{Z_j, \colon j=1,\ldots,i-1,i+1,\ldots,n,n+1\}$. For NCMs that incorporate information from the entire set $\mathcal{Z}_{(-i)}$, like the one we are proposing in Section \ref{efdm}, the process of retraining for every $i\in\{1,\ldots,n,n+1\}$ can become computationally expensive.

\emph{Inductive} CP (ICP) is an alternative to transductive CP that has a lower computational burden at the cost of less efficient use of available data. ICP randomly splits all available observations for training into two sets: a \emph{training data set} and a \emph{calibration data set}. The purpose of the training data set is to fit a base model which relates the object $X$ to the label $Y$. Meanwhile, the purpose of the calibration data set is to evaluate the NCM. For clarity, we use the term \emph{full training data set} to refer to all observations available for training and denote this set as $D^{full} = \{Z_1,\ldots,Z_n\}$. Similarly, we denote the training and calibration sets, respectively, as $D^{tr} = \{Z_1, \ldots, Z_{n_1}\}$, and $D^{cal} = \{Z_{n_1+1}, \ldots, Z_{n_1 + n_2}\}$. Let $\mathcal{I}^{tr} = \{1,\ldots,n_1\}$ and $\mathcal{I}^{cal} = \{n_1+1, \ldots, n_1+n_2\}$ represent the indices of the observations in $D^{tr}$ and $D^{cal}$, respectively. As is commonly used \citep{vovk2015}, we take $n_1 = \left\lceil \frac{2}{3} n\right\rceil$ and $n_2 = n - n_1$ where $\lceil \cdot \rceil$ is the ceiling function. To construct a prediction set for a test observation, $Z_{n+1}$, the NCM is computed and compared to the NCM values obtained by evaluating the NCM on the calibration data set. 

In this work, we use the p-value approach to constructing prediction sets. Let $s_{n_1+i} = s(Z_{n_1+i};D^{tr})$ represent the NCM value of the $i^{th}$ calibration observation and $s_{n+1}^y= s((X_{n+1},y);D^{tr})$ be the NCM value of the test point where $y$ is the assumed value of $Y_{n+1}$, the label associated with $X_{n+1}$. The p-value corresponding to $X_{n+1}$ is then computed as
\begin{align} \label{p_val_form}
    p_{n+1}^y &= \frac{|\{i \in \mathcal{I}^{cal} : s_i \ge s_{n+1}^y\}|+1}{n_2 + 1}
\end{align}
In traditional CP, when $p_{n+1}^y\ge\alpha$, the assumed value, $y$, of the label $Y_{n+1}$, is one element that defines the prediction set, $\mathcal{C}(X_{n_1})$. In the case of ICAD, there is no reliance upon the label, and to simplify the notation we drop the superscript $y$ and denote the p-value as $p_{n+1}$. If exchangeability assumptions hold and $p_{n+1} \ge \alpha$ then $Z_{n+1}$ is labeled as an inlier with $(1-\alpha)\cdot100\%$ confidence. Otherwise, $Z_{n+1}$ is labeled an outlier. Note that the marginal coverage guarantee is only applicable to inliers and not to outliers. Finally, we must use $\alpha \ge \frac{1}{n_2+1}$ or all test points will automatically be labeled as inliers.

\section{Elastic Functional Distance Metrics ICAD} \label{efdm}
Our proposed method is ICAD where the NCM is based on elastic functional distances and the Karcher mean. We call this method \emph{elastic functional distance metrics} ICAD, or EFDM. Let $D^{full} = \{f_{(1)}, \ldots, f_{(n)}\}$ be the full training data set of functional data observations. Further assume $f_1, \ldots, f_n \sim \mathcal{P_F}$ are exchangeable, where $\mathcal{P_F}$ is a probability distribution over the function space $\mathcal{F}$. Using random assignment we create $D^{tr} = \{f_1, \ldots, f_{n_1}\}$ and $D^{cal} = \{f_{n_1+1}, \ldots, f_{n_1 + n_2}\}$. Due to the random assignment, the indices in $D^{tr}$ and $D^{cal}$ are different than those in $D^{full}$. To keep these clear, $f_{(i)}$ refers to the $i^{th}$ functional observation in the full training data while $f_i$ and $f_{n_1+i}$ refer to the $i^{th}$ functional observations in the training and calibration sets, respectively. As described in Section \ref{intro}, we also assume that all functional observations (training, calibration, and testing observations), are measured at the same points in the domain. We denote these domain points as $\mathcal{T} = \{t_0, t_1, \ldots, t_M\}$.

To label a new function $f_{n+1}$ as an inlier or outlier using a level of significance $\alpha$, EFDM proceeds as follows:
\begin{enumerate}
    \item Compute the Karcher mean of the training data set $D^{tr}$ using Equation \ref{eq:sec2_kmf} (or equations \ref{eq:sec2_kmq} and \ref{eq:sec2_reconstructfromq}), denote the Karcher mean as $\mu_{tr}$.

    \item Compute $d_{a\,i}^{tr} = d_a(\mu_{tr}, f_i)$ and $d_{p\,i}^{tr} = d_p(\mu_{tr}, f_i)$, the amplitude and phase distances from the Karcher mean to each function in the training data. Let $min_a$ and $max_a$ represent the minimum and maximum amplitude distances, respectively; and $min_p$ and $max_p$ similarly for the phase distances.

    \item Compute $d_{a\,i}^{cal} = d_a(\mu_{tr}, f_{n_1+i})$ and $d_{p\,i}^{cal} = d_p(\mu_{tr}, f_{n_1+i})$, the amplitude and phase distances from the Karcher mean to each function in the calibration data.

    \item Compute the NCM for each calibration observation as
    $$s_i = \frac{1}{2}\left[ \left(\frac{d_{a\,i}^{cal} - min_a}{max_a - min_a}\right) + \left(\frac{d_{p\,i}^{cal} - min_p}{max_p - min_p}\right) \right]$$

    \item Compute $d_a^{ts} = d_a(\mu_{tr}, f_{n+1})$ and $d_p^{ts} = d_p(\mu_{tr}, f_{n+1})$, the amplitude and phase distances from the Karcher mean to the test function.

    \item Compute the NCM for the test function as 
    $$s_{n+1} = \frac{1}{2}\left[ \left(\frac{d_{a}^{ts} - min_a}{max_a - min_a}\right) + \left(\frac{d_{p}^{ts} - min_p}{max_p - min_p}\right) \right]$$

    \item Compute the p-value, $p_{n+1}$, as in equation (\ref{p_val_form}). If $p_{n+1} < \alpha$, $f_{n+1}$ is labeled as an outlier; else it is labeled as an inlier.
\end{enumerate}
As seen in steps 4 and 6, the NCM used for EFDM is the average of the scaled amplitude and phase distances from an observation to the Karcher mean of the training data. This scaling is important so that both the amplitude and phase components are unitless and can contribute equally to the NCM. We now describe several potential adjustments to this NCM.

The basic NCM given above is primarily intended to detect shape outliers. This approach will also detect magnitude outliers if they are outlying \emph{on the $x$-axis} (phase distance will properly account for these). However, magnitude outliers that are outlying predominately in the $y$ direction may not be accounted for by amplitude distance (this is because the functions are first transformed to SRSFs before computing amplitude distance). In short, magnitude outliers that are essentially vertical shifts of the data will not be detected. To enable the detection of such magnitude outliers, \emph{translation distance} can be incorporated into the NCM. To do this, in step 2, we compute $|\mu_{tr}(t_0) - f_i(t_0)|$ for $i = 1, \ldots, n_1$ along with minimum and maximum values for scaling. In step 3, we compute $|\mu_{tr}(t_0) - f_{n_1+i}(t_0)|$ for $i = 1, \ldots, n_2$. In step 4, the multiplier becomes $\frac{1}{3}$ instead of $\frac{1}{2}$, and a term for the scaled translation distance is included within the brackets. The same is also done for the test point, $f_{n+1}$.

Another possible adjustment is to allow for different weightings of the amplitude and phase (and potentially translation) distance components so that the NCM is a weighted average rather than a simple arithmetic mean. While a functional data set that displays more amplitude than phase variability, or vice versa, is not a problem because we scale the distances, there may still be cases when it is desirable to weight one component higher than the other. Taking this to an extreme, it is also possible to use only one of the components if, for instance, we know \emph{a prioi} that there is only one type of variability in a data set. We caution against this, however. Suppose a functional data set that contains only amplitude variability is used in the EFDM algorithm with only amplitude distance in the NCM. If a new observation looks similar to the training data in amplitude but varies in phase, it will not be marked as an outlier.

Finally, smoothed conformal prediction \citep{vovk2005} is often used to guarantee exact asymptotic validity. This is carried out by adjusting the p-value computation to 
\begin{align}
    \Tilde{p}_{n+1} = \frac{|\{i \in \mathcal{I}^{cal}: s_i > s_{n+1}\}|+\tau |\{i \in \mathcal{I}^{cal} \cup \{n+1\} : s_i = s_{n+1}\}|}{n_2 + 1}
\end{align}
where $\tau$ is a random draw from the $U(0,1)$ distribution. In all of our computations, we use these smoothed conformal p-values. Other p-value adjustments are possible and may be desirable, depending on the use case (see, e.g., \citealp{bates2023}).

\subsection{Other Functional ICAD Methods} \label{efdm:gmd_sncm}
In this section, we frame the methods of \cite{lei2015} and \cite{diquigiovanni2021} as functional ICAD methods for comparison to EFDM.

\subsubsection{GMD} \label{efdm:gmd_sncm:gmd}
The GMD ICAD method begins by projecting the functional data to a lower-dimensional space. In our case, we use functional principal component analysis (FPCA) \citep{Ramsay2005} to achieve the projection, but other projection methods could be used. Let $\xi_{ij} = \langle f_i, \theta_j \rangle$ be the $j^{th}$ component of the projected functional observation, $f_i$ for $i = 1, \ldots, n$ and $j = 1, \ldots, p$. Here, $\theta_j$ represents the $j^{th}$ functional principal component and $\langle \cdot, \cdot \rangle$ is the functional inner product \citep{Ramsay2005}. The projection of $f_i$ can then be written as $\xi_i = (\xi_{i1}, \ldots, \xi_{ip})$. Note that the FPCs are estimated using only the training data, and all training, calibration, and test data use the same FPCs for projection to $\xi_i$.

The projected data are modeled with a Gaussian mixture model (GMM) \citep{hastie2009} with $K$ components. From the projected version of the training data, $D_{\xi}^{tr} = \{\xi_1, \ldots, \xi_{n_1}\}$, the mean, covariance and mixture proportions are learned. We denote the fitted GMM as 
$$G(\xi) = \sum_{k=1}^K \hat{\pi}_k \phi(\xi; \hat{\mu}_k, \hat{\Sigma}_k)$$
where $\phi(\cdot; \mu, \Sigma)$ represents the multivariate normal density with mean vector $\mu$ and covariance matrix $\Sigma$, and the $\hat{\pi}_k$ are the estimated mixing proportions. The NCM is then computed as $s_i(\xi_i) = -1\cdot G(\xi_i)$, hence the name Gaussian mixture density (GMD) for this approach\footnote{Note that \cite{lei2015} use a \emph{conformity} measure rather than a non-conformity measure, so they simply compute the density rather than the negative density. The two approaches are equivalent.}. We also implement a refinement that was introduced by \cite{lei2015} to provide a narrower prediction band. With the GMD maximum (GMDM) refinement, the NCM is instead computed as 
$$s_i(\xi_i) = -1\cdot \left(\max_{\{1,\ldots, K\}} \hat{\pi}_k \phi(\xi_i; \hat{\mu}_k, \hat{\Sigma}_k)\right).$$

We wrote our own implementations of GMD and GMDM in \texttt{R}. Our code uses the \texttt{mclust} package (version 6.1.1) \citep{scrucca2023} and the \texttt{mvtnorm} package (version 1.3.1) \citep{genz2021}.

\subsubsection{SNCM} \label{efdm:gmd_sncm:sncm}
In the work of \cite{diquigiovanni2021}, the proposed NCM is computed as 
$$s_i(f_i) = \sup_{t \in \mathcal{T}}\left|\frac{f_i(t) - m(t)}{r(t)}\right|$$
where $m(t)$ is a mean function and $r(t)$ is a modulation function, and both of these are estimated from only the training data. Because this NCM uses the supremum, we refer to this method as supremum non-conformity measure (SNCM) ICAD. For the mean function, only the cross-sectional mean function is used by \cite{diquigiovanni2021}. For the modulation function, the authors compare three options: $r(t) = 1$; the cross-sectional standard deviation function; and what is referred to as the optimal modulation function. The optimal modulation function is defined as
$$r(t) = \frac{\max_{i \in H_1}|f_i(t) - m(t)|}{\int_{\mathcal{T}}\max_{i \in H_1}|f_i(t) - m(t)|dt}$$
where 
\begin{equation*}
    H_1 = \begin{cases}
        \mathcal{I}^{tr}&\text{if }\lceil (n_1 + 1)(1-\alpha) \rceil > n_1\\
        \{i\in\mathcal{I}^{tr}\colon\sum_{t\in\mathcal{T}}\vert f_i(t)-m(t)\vert\leq\nu\}&\text{if }\lceil (n_1 + 1)(1-\alpha) \rceil \le n_1
    \end{cases}
\end{equation*}%
where $\nu$ is the $\lceil (n_1 + 1)(1-\alpha) \rceil^{th}$ smallest value of the set $\{\sup_{t \in \mathcal{T}} |f_i(t) - m(t)| : i \in D^{tr}\}$.

We wrote our own implementation of SNCM in \texttt{R}. Our code made use of the \texttt{fdasrvf} package (version 2.3.4) and the \texttt{caTools} package (version 1.18.3) \citep{tuszynski2024}.

\subsection{Leave-one-out Approach} \label{efdm:loo}
Because CP assumes exchangeability, all ICAD methods are best suited for the first outlier detection scenario where a pristine data set is used for training and calibration, and new or external data are labeled by computing conformal p-values. However, we also propose a leave-one-out approach to address the second outlier detection scenario where only a single data set is available and a user wishes to know which observations are the most anomalous. If outliers exist in a data set, the data are no longer exchangeable. In such a case, we lose the coverage guarantee provided by ICAD, but we believe the basic approach can still be useful for detecting outliers.

In our proposed approach, we need to assign inlier/outlier status to every $f_{(i)} \in D^{full}$. To do so, we let $D_{-(i)}^{full}$ be the full training data set with $f_{(i)}$ excluded. We then randomly split $D_{-(i)}^{full}$ into training and calibration sets and proceed with ICAD as before to produce a p-value and outlier determination for $f_{(i)}$. This process is repeated until p-values have been obtained for all $n$ functional observations.

\section{Simulation Experiments} \label{sim}
In this section, we empirically evaluate the performance of the methods described in Section \ref{efdm}. Magnitude outliers, as mentioned in Section \ref{intro:litrev:fnout}, are the easier case, so these experiments focus instead on detecting shape outliers.

\subsection{Experiment Data} \label{sim:setup}
These experiments use functional data generated from three different templates. These are the \emph{standard}, \emph{narrow}, and \emph{double peaked} templates. Respectively, they are given as:
$$F_{st}(t) = a\exp\left[-\frac{1}{2}(t-p)^2\right]$$
$$F_{nr}(t) = a\exp\left[-2(t-p)^2\right]$$
$$F_{dp}(t) = a\exp\left[-\frac{1}{2}(t - 1.5 - p)^2\right] + a\exp\left[-\frac{1}{2}(t + 1.5 - p)^2\right].$$

\begin{figure}
    \centering
    \begin{tabular}{ccc}
    \includegraphics[width=0.33\linewidth]{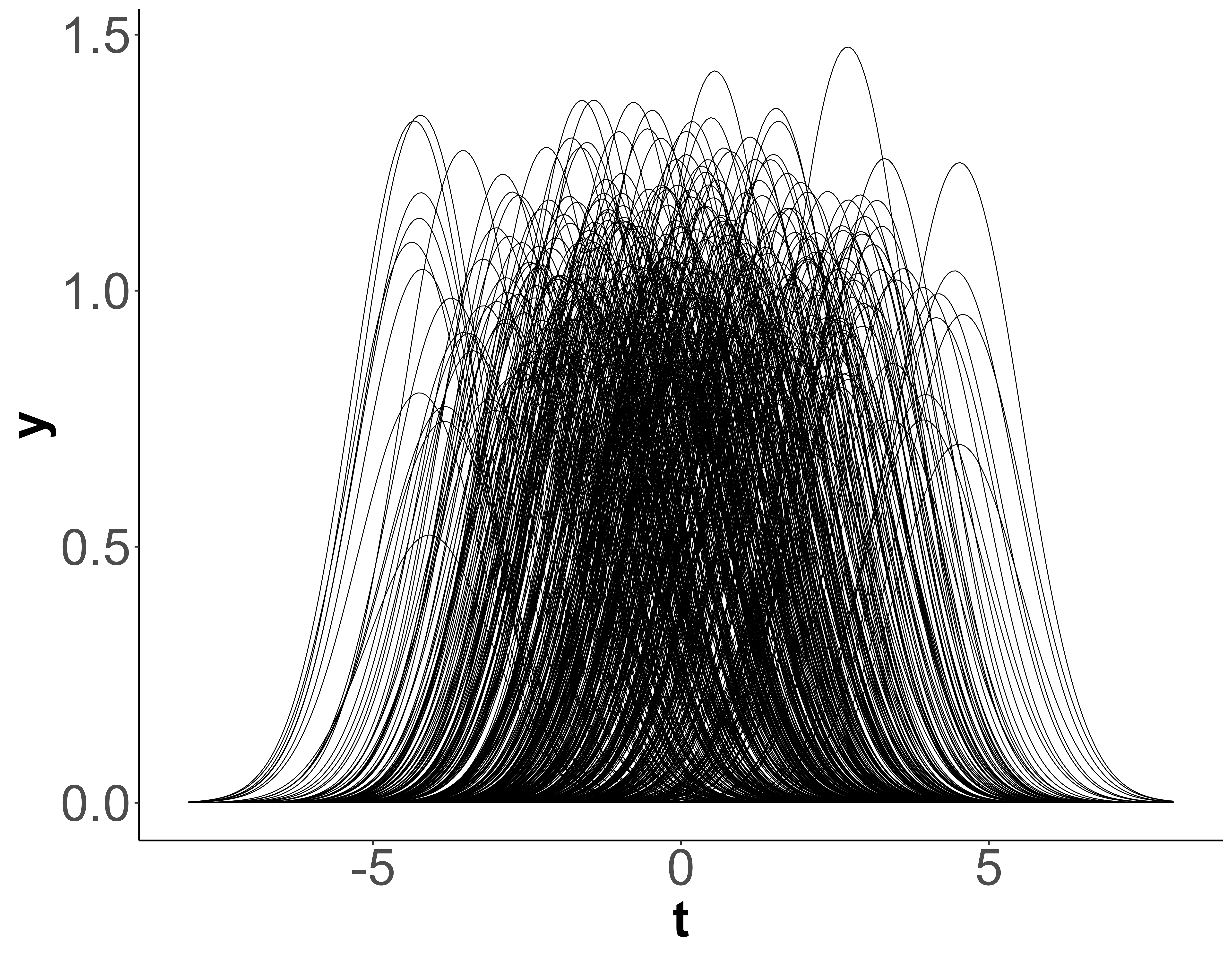} &  
    \includegraphics[width=0.33\linewidth]{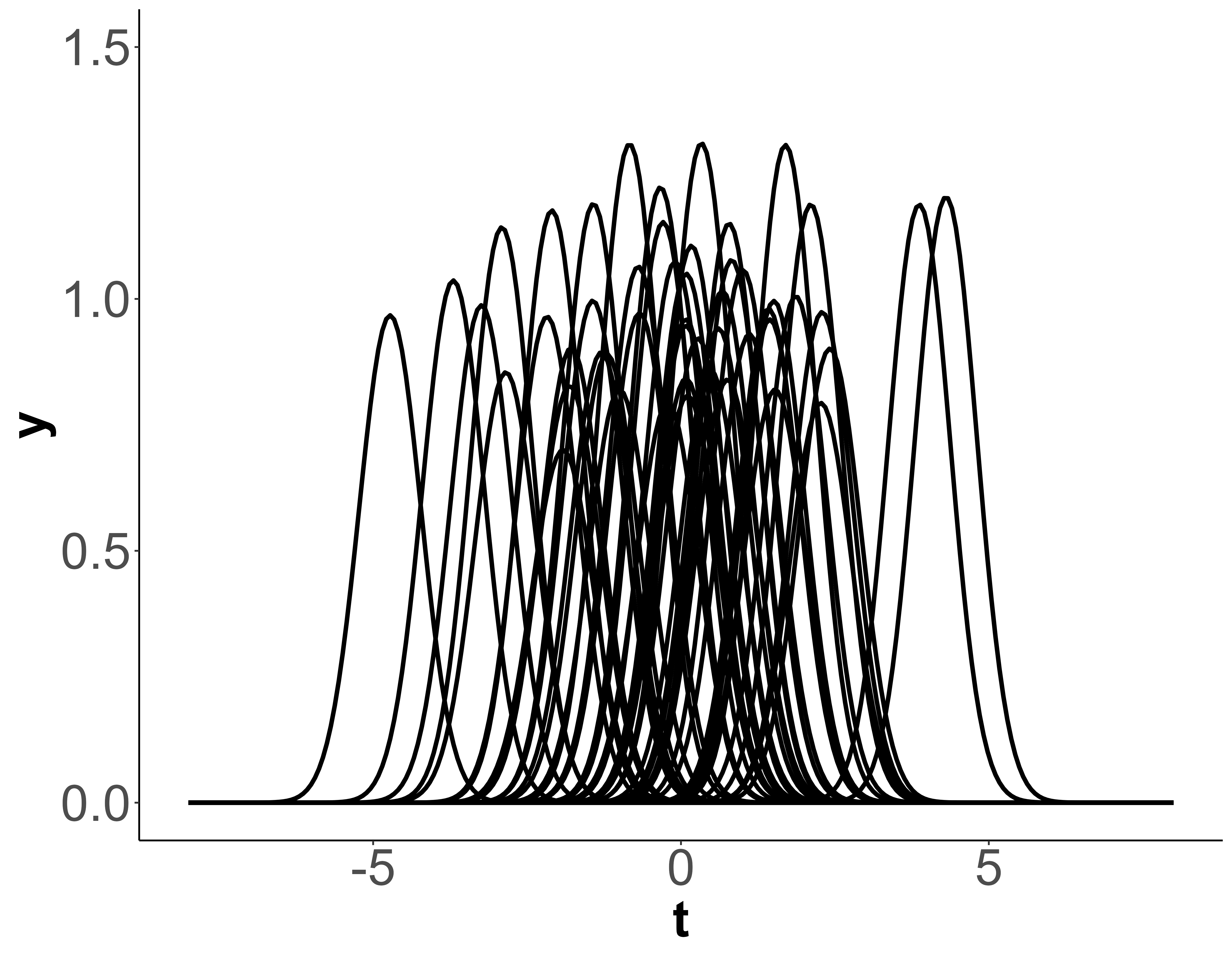} & 
    \includegraphics[width = 0.33\linewidth]{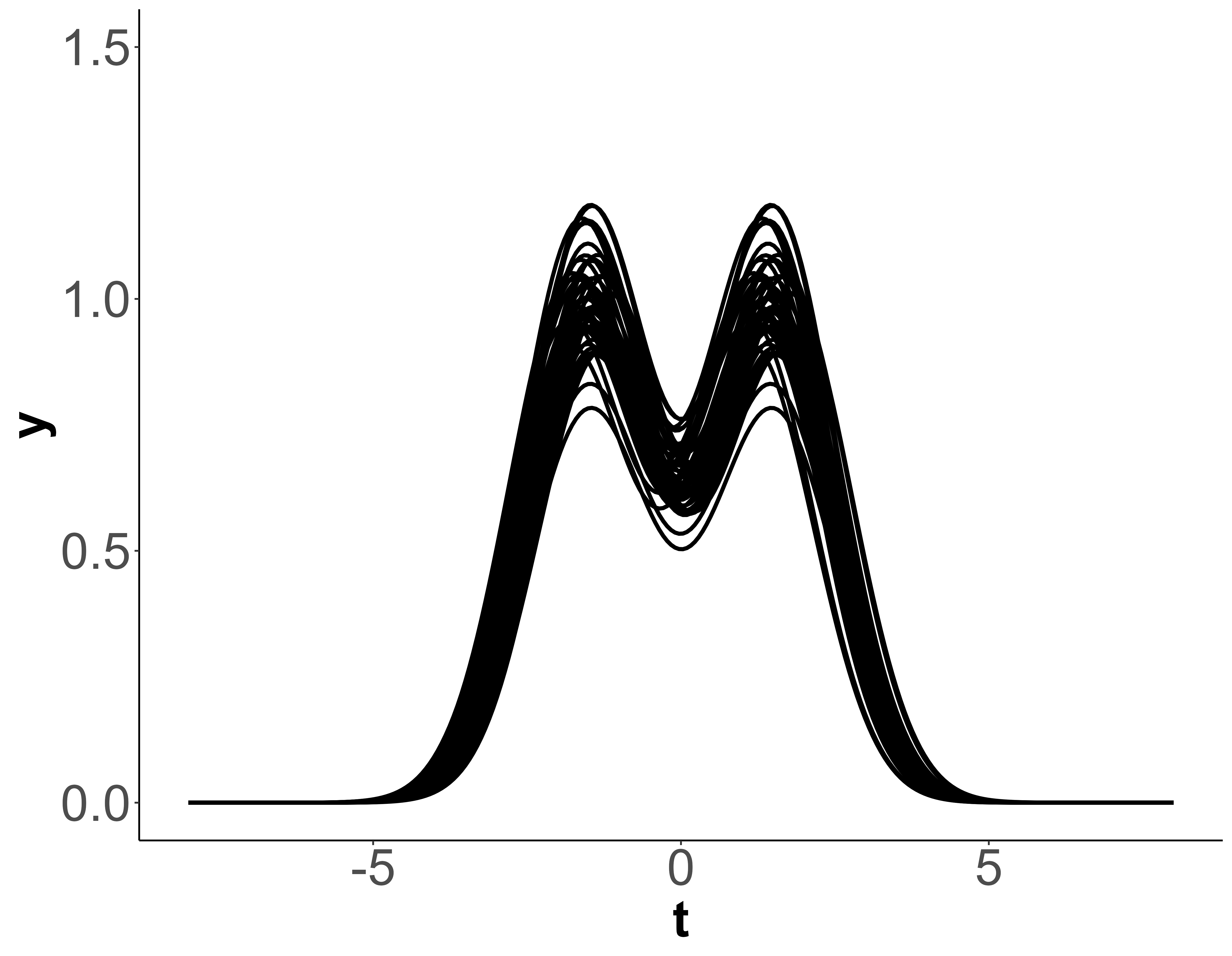} \\
    (a) & (b) & (c)
    \end{tabular}
    \caption{Randomly generated data sets. (a) 500 standard functions, (b) 50 narrow functions, and (c) 50 double peaked functions. These are also the test sets for the first experiment.}
    \label{fig:sim_dat}
\end{figure}

To generate functional data from these templates, we randomly generate values for the $a$ and $p$ parameters. Figure \ref{fig:sim_dat} shows randomly generated data from each of these templates. Panel (a) contains 500 standard functions, panel (b) contains 50 narrow functions, and panel (c) contains 50 double peaked functions. The standard and narrow data sets were generated using $a \sim N(\mu = 1, \sigma = 0.15)$ and $p \sim N(0, 1.75)$. The double peaked functions were generated with $a \sim N(1, 0.10)$ and $p \sim N(0, 0.15)$. The domain points for all generated functions are 250 equally spaced points over the interval $[-8,8]$. Treating the standard data as inliers, the key thing to note about both the narrow and double peaked data is that these are \emph{shape-only} outliers. In terms of magnitude, they fit well within the distribution of the standard data.

\subsection{Experiment 1} \label{sim:exp1}
The goal of the first experiment is two-fold: first, we investigate the coverage of the ICAD methods over inlier and outlier functions separately. Second, we consider the effect of sample size on the coverage. Using full training data set sizes of $n = 50$, $100$, and $250$, we randomly generate 500 data sets using the standard template. That is, there are 1500 total simulated full training data sets for this experiment. Each of the 1500 data sets are run through the ICAD methods and labels are assigned to each function in three different test sets. The first test set consists of 500 standard functions generated with the same parameters as the full training sets (these are the inlier functions). The second test set contains 50 narrow functions, and the third contains 50 double peaked functions. Both the second and third test sets are outliers relative to the training data. These test sets are shown in Figure \ref{fig:sim_dat}. 

\begin{table}
    \centering
    \resizebox{\textwidth}{!}{
    \begin{tabular}{r|l} 
        Method Name & Details \\ \hline
        EFDM & NCM uses amplitude and phase distances only \\
        SNCM1 & Uses the cross-sectional mean function and the optimal modulation function \\
        SNCM2 & Uses the Karcher mean and the optimal modulation function \\
        GMD1 & Uses $p = 2$, $K = 1$ \\
        GMD2 & Uses $p=3$, $K = 2$ \\
        GMD3 & Uses $p = 5$, $K = 2$ \\
        GMDM & Uses $p=5$, $K = 2$ 
    \end{tabular}}
    \caption{Description of methods used for experiment 1.}
    \label{tab:exp1_methods}
\end{table}

Table \ref{tab:exp1_methods} details each method used for this experiment. For each simulated data set, inlier/outlier labels were obtained for every functional observation in the three test sets using a significance level of $\alpha = 0.10$. The coverage for a single simulated data set is simply the proportion of test functions marked as inliers (this can be viewed as the coverage averaged over the test functions). These proportions are then averaged over all 500 simulated data sets that have the same sample size to provide the final mean coverage estimate for a method. These values, along with the standard deviation over the simulated data sets, are given in Tables \ref{tab:exp1_st_covg} and \ref{tab:exp1_nrdp_covg}. Table \ref{tab:exp1_st_covg} contains results from the standard test set where the mean coverage values should be close to $1 - \alpha = 0.90$. Table \ref{tab:exp1_nrdp_covg} contains results from the narrow and double peaked test sets where mean coverage values close to $0$ are desirable. Note that the values for GMD3 and GMDM are missing because, with some of the full training sets of size $n=50$ and $n=100$, the GMMs could not be fit. In all tables, bolded values represent the best value(s) in each column.

\begin{table}[htbp]
    \centering    
    \begin{tabular}{|c|c|c|c|} \hline
     & \multicolumn{3}{|c|}{Standard Test Data} \\ \hline
     Method & $n=50$ & $n=100$ & $n=250$  \\ \hline
     EFDM & $0.893 (0.07)$ & $0.895 (0.05)$ & $\textbf{0.901 (0.03)}$   \\ \hline
     SNCM1 & $\textbf{0.900 (0.06)}$ & $\textbf{0.899 (0.05)}$ & $\textbf{0.899 (0.03)}$  \\ \hline
     SNCM2 & $0.901 (0.06)$ & $\textbf{0.899 (0.05)}$ & $0.903 (0.03)$  \\ \hline
     GMD1 & $0.901 (0.07)$ & $\textbf{0.899 (0.05)}$ & $0.905 (0.04)$   \\ \hline
     GMD2 & $\textbf{0.900 (0.06)}$ & $\textbf{0.899 (0.05)}$ & $\textbf{0.901 (0.03)}$   \\ \hline
     GMD3 & $-$ & $-$ & $\textbf{0.901 (0.03)}$  \\ \hline
     GMDM & $-$ & $-$ & $0.884 (0.03)$  \\ \hline
    \end{tabular}
    \caption{Mean(SD) coverage results from experiment 1 for the standard test data.}
    \label{tab:exp1_st_covg}
\end{table}

\begin{figure}[htbp]
    \centering
    \begin{tabular}{cc}
      \includegraphics[width=0.50\linewidth]{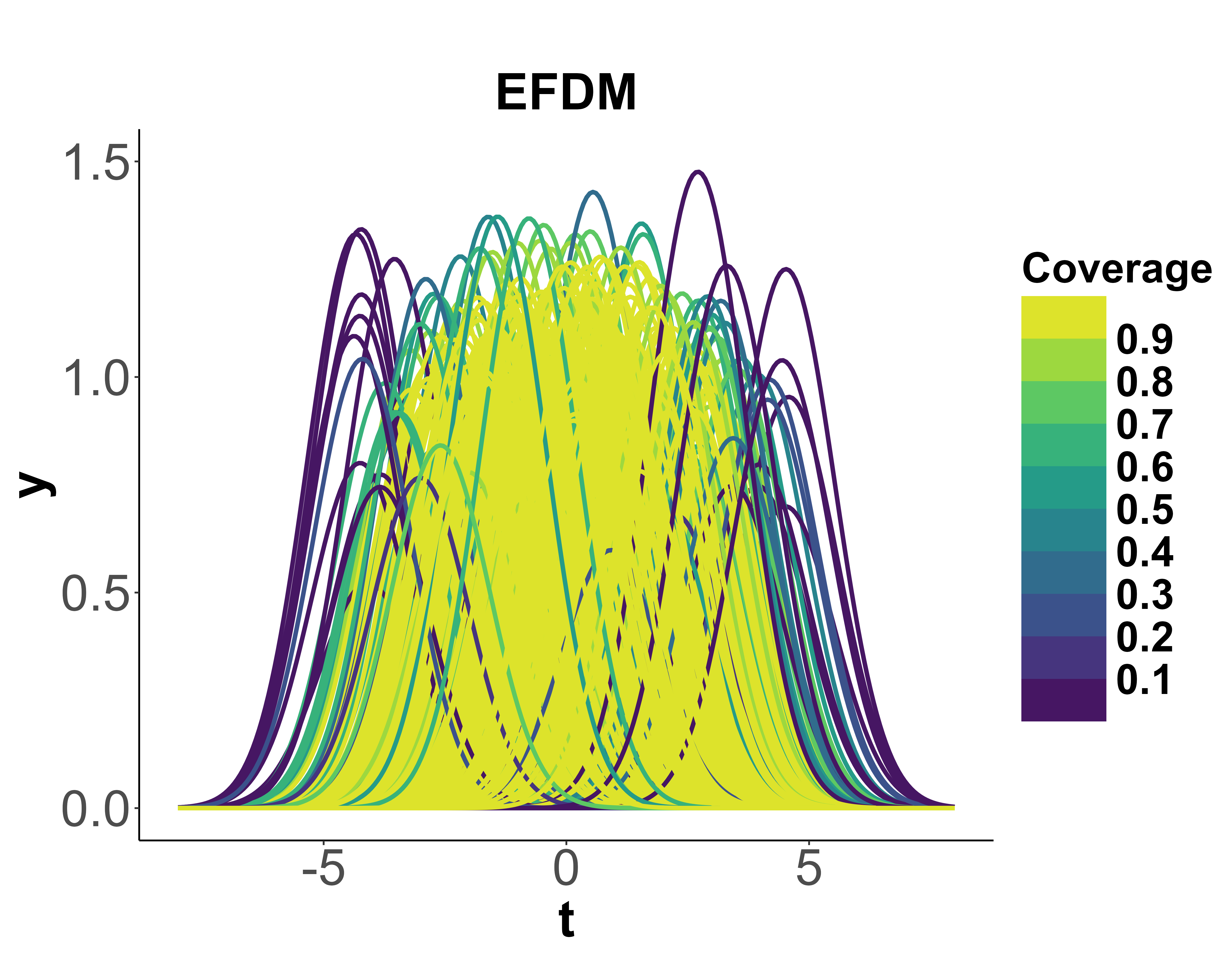} &  
      \includegraphics[width=0.50\linewidth]{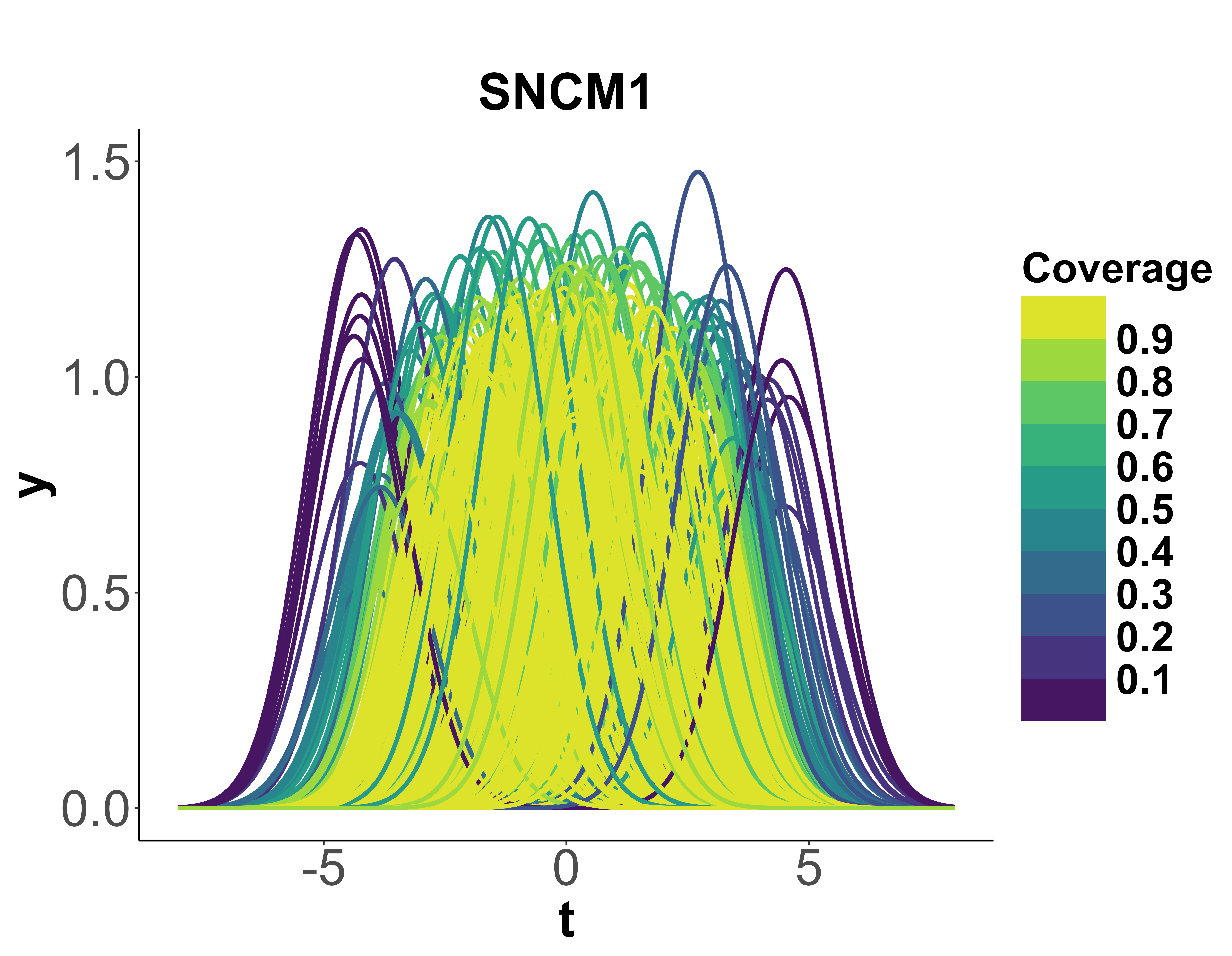} \\
      \includegraphics[width=0.50\linewidth]{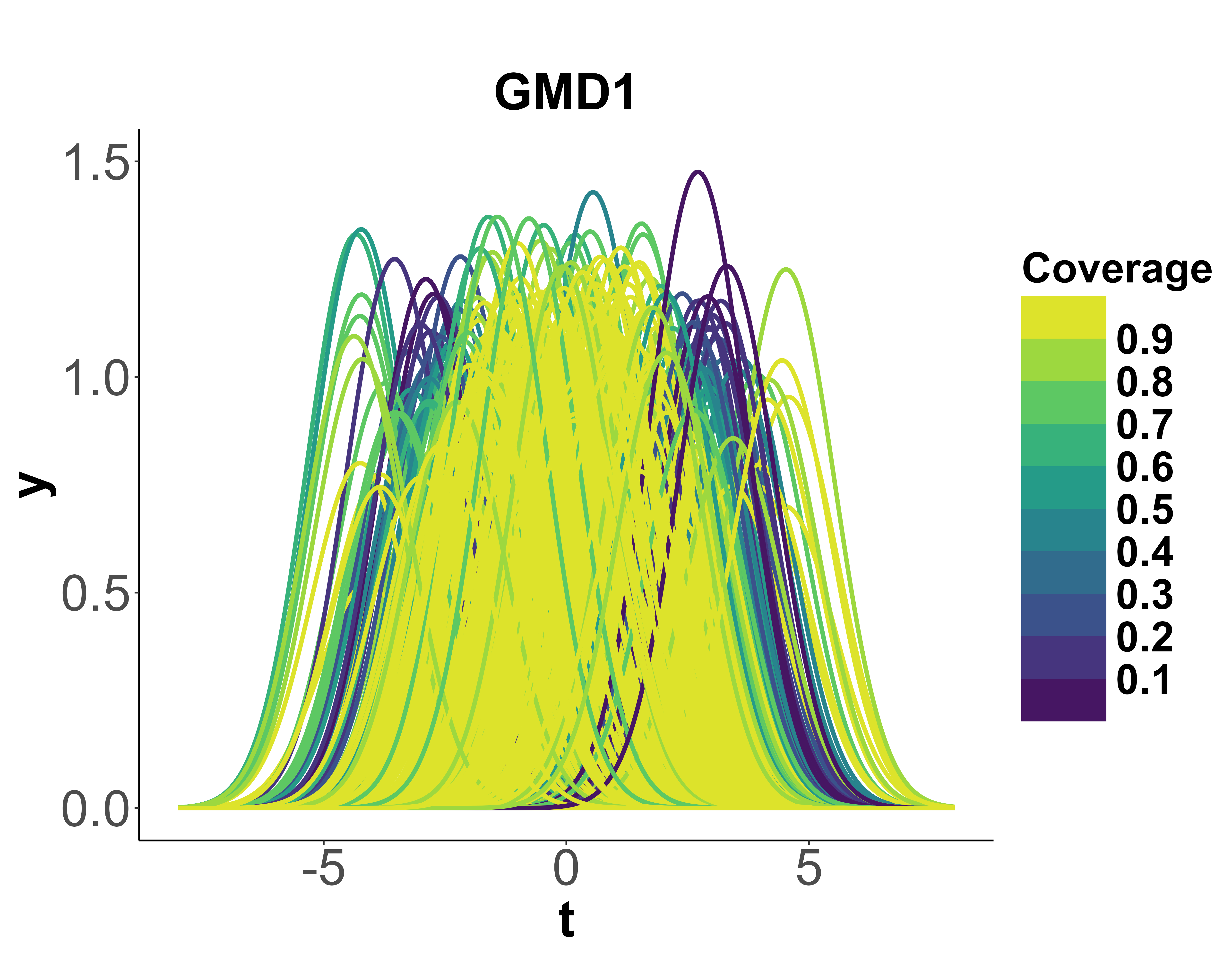} &
      \includegraphics[width=0.50\linewidth]{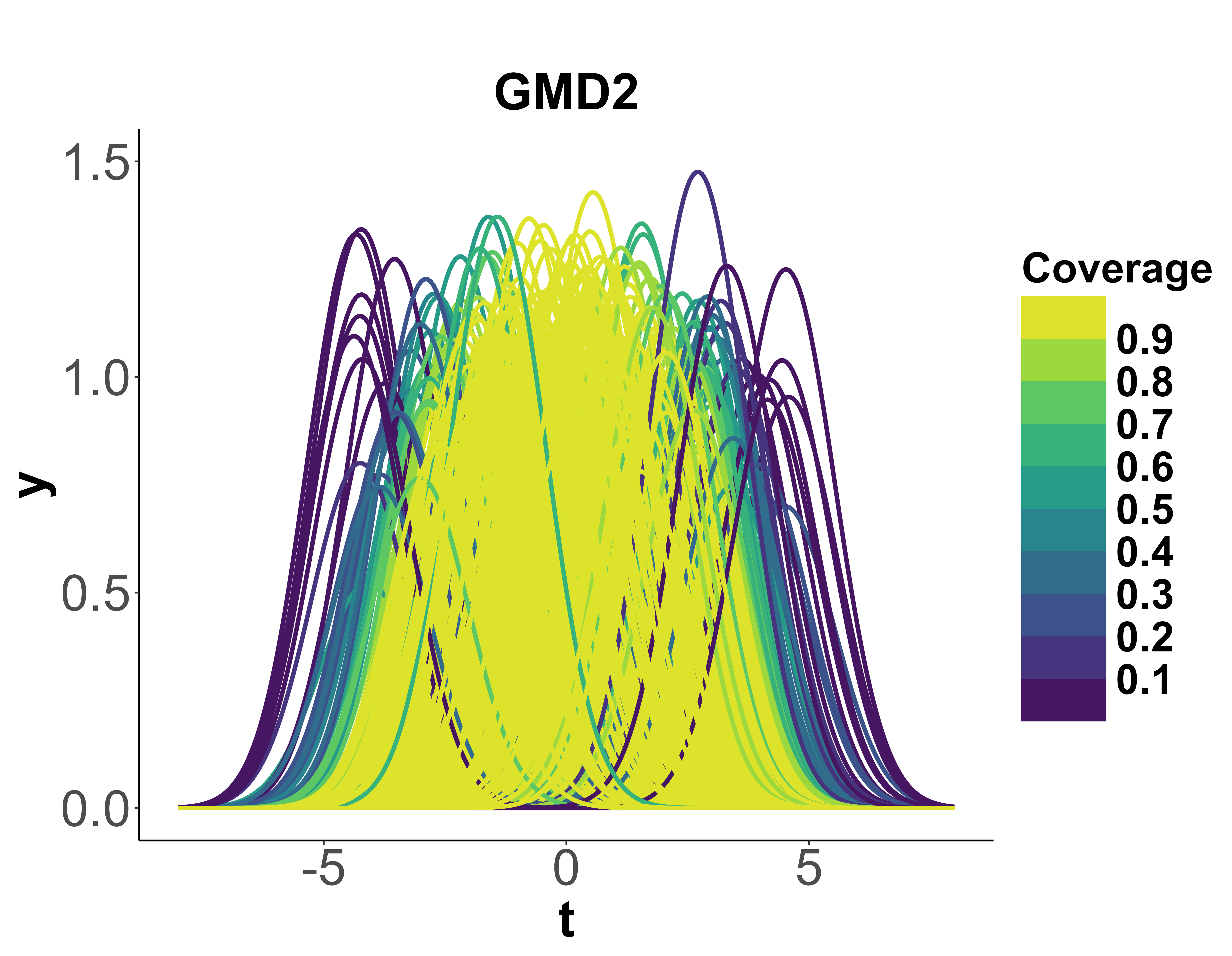} \\
    \end{tabular}
    \caption{Plots of average coverage per test function in the standard test data for EFDM, SNCM1, GMD1, and GMD2 when $n=100$.}
    \label{fig:exp1_covg}
\end{figure}

Figure \ref{fig:exp1_covg} contains plots of the test data sets where the color of each function represents the proportion of times it is covered over the 500 replications. While separate plots could be given for every filled cell in Tables \ref{tab:exp1_st_covg} and \ref{tab:exp1_nrdp_covg}, we only show those corresponding to the standard test data for EFDM, SNCM1, GMD1, and GMD2 where $n=100$. Plots corresponding to the narrow and double peaked test data for these same methods when $n=100$ are given in Appendix \ref{appdx}.

\begin{table}[htbp]
    \centering
    \resizebox{\textwidth}{!}{
    \begin{tabular}{|c|c|c|c|c|c|c|} \hline
     & \multicolumn{3}{|c|}{Narrow Test Data} &\multicolumn{3}{|c|}{Double Peaked Test Data}   \\ \hline
     Method & $n=50$ & $n=100$ & $n=250$ & $n=50$ & $n=100$ & $n=250$ \\ \hline
     EFDM &  $\textbf{0.052(0.13)}$ & $\textbf{0.001(0.02)}$ & $\textbf{0(0)}$ & $\textbf{0.001(0.01)}$ & 
     $\textbf{0(0)}$ & $\textbf{0(0)}$ \\ \hline
     SNCM1 &  $0.927 (0.08)$ & $0.911 (0.06)$ & $0.888 (0.04)$ & $0.972 (0.08)$ & $0.982 (0.04)$ & $0.991 (0.03)$ \\ \hline
     SNCM2 &  $0.927 (0.09)$ & $0.908 (0.07)$  & $0.882 (0.08)$ & $0.966 (0.08)$ & $0.978 (0.05)$ & $0.992 (0.03)$ \\ \hline
     GMD1 &  $0.992 (0.02)$ & $0.997 (0.01)$ & $0.999 (0.002)$ & $0.999 (0.002)$ & $1 (0)$ & $1 (0)$ \\ \hline
     GMD2 &  $0.889 (0.13)$ & $0.900 (0.09)$ & $0.912 (0.04)$ & $0.389 (0.34)$ & $0.241 (0.26)$ & $0.131 (0.17)$ \\ \hline
     GMD3 &  $-$ & $-$ & $0.542 (0.19)$ & $-$ & $-$ & $0.170 (0.21)$ \\ \hline
     GMDM &  $-$ & $-$ & $0.487 (0.19)$ & $-$ & $-$ & $0.104 (0.15)$ \\ \hline
    \end{tabular}}
    \caption{Mean(SD) coverage results from experiment 1 for the narrow and double peaked test data.}
    \label{tab:exp1_nrdp_covg}

\end{table}

Because the standard test data and the training data sets are generated from the same process, we expect the assumptions of ICAD to hold. This is borne out in Table \ref{tab:exp1_st_covg} where mean coverage values are all very close to $0.90$ for all methods. The only effect of sample size seems to be to reduce the variability in the coverage estimates for each data set, as would also be expected. From this table, no methods stand out as outperforming the others. The plots in Figure \ref{fig:exp1_covg} are somewhat more informative as they demonstrate how each method attains the correct average coverage. The EFDM method appears to be the only one that identifies functions with small maximum values in the middle of the data as not belonging to the distribution. For GMD1, it is somewhat strange that the most extreme functions in terms of phase distance are labeled as inliers more often than less extreme functions. GMD2 does not exhibit this behavior.

The performance of the methods is much more variable in Table \ref{tab:exp1_nrdp_covg}. Here, EFDM has the best performance in every case. With enough training data ($n=250$ for the narrow data and $n=100$ for the double peaked), it never labels outliers as inliers in any of the 500 runs. Both SNCM methods perform poorly, but the coverage does decrease somewhat as sample size increases for the narrow test data. This pattern, however, does not hold for the double peaked data. GMD1 has the worst performance of any method in every case. GMD2 performs poorly on the narrow test data but much better on the double peaked data. GMD3 and GMDM perform better on the narrow test data than the other GMD and SNCM methods, and they even outperform GMD2 on the double peaked data.

\subsection{Experiment 2} \label{sim:exp2}
In experiment 2, our goal is to simulate the first outlier detection scenario where the training and calibration data contain no outliers and external data, which may contain outliers, are labeled using the ICAD methods. In this experiment, we reuse the same 1500 full training data sets from experiment 1 (so there are 500 data sets for each sample size, $n=50$, $n=100$, and $n=250$). Each full training set has two corresponding test sets containing 100 functional observations each. One test set contains $95\%$ inliers (i.e., they are simulated using the standard template function and the same normal distribution parameters as the full training data) and $5\%$ double peaked outliers, simulated with the same parameters as the double peaked test data in experiment 1. The other test set contains $90\%$ inliers and $10\%$ double peaked outliers.

Each full training data set is randomly split into training and calibration data, and each of the methods from Table \ref{tab:exp1_methods} are used to assign inlier/outlier status to all functions in both test data sets. For each test data set, we examine the effects of using two different levels of significance, $\alpha = 0.05$ and $\alpha = 0.10$. Method performance is assessed using the Matthew's correlation coefficient (MCC) \citep{matthews1975}, which has been shown to perform well in cases of extreme class imbalance \citep{chicco2021}. MCC values close to 1 indicate better performance (or higher positive correlation between predictions and ground truth labels) while values close to 0 indicate poor performance (little correlation between predictions and ground turth labels). Since it is a correlation, MCC can also be negative (indicating negative correlation between predictions and ground truth labels). For this experiment, the double peaked outliers are treated as the positive class. MCC is computed for each test set, and the mean and standard deviation over the 500 full training sets (per sample size) are given in Tables \ref{tab:exp2_mcc05} and \ref{tab:exp2_mcc10}. As in experiment 1, values are missing for GMD3 and GMDM when $n=50$ and $n=100$. Additional metrics from this experiment are given in Appendix \ref{appdx}.

\begin{table}[htbp]
    \centering
    \Large
    \resizebox{\textwidth}{!}{\begin{tabular}{|c|c|c|c|c|c|c|} \hline
      & \multicolumn{3}{|c|}{$5\%$ Outliers in Test Data}   & \multicolumn{3}{|c|}{$10\%$ Outliers in Test Data} \\ \hline
      Method & $n=50$ & $n=100$ & $n=250$ & $n=50$ & $n=100$ & $n=250$ \\ \hline
      EFDM & $\textbf{0.668(0.19)}$ & $\textbf{0.734(0.14)}$ & $\textbf{0.728(0.12)}$ & $\textbf{0.735(0.15)}$ & $\textbf{0.818(0.11)}$ & $\textbf{0.817(0.10)}$ \\ \hline
      SNCM1 & $-0.033(0.05)$ & $-0.037(0.04)$ & $-0.044(0.03)$ & $-0.049(0.06)$ & $-0.058(0.04)$ & $-0.063(0.04)$ \\ \hline
      SNCM2 & $-0.027(0.06)$ & $-0.034(0.05)$ & $-0.044(0.03)$ & $-0.042(0.07)$ & $-0.053(0.05)$ & $-0.062(0.04)$ \\ \hline
      GMD1 & $-0.043(0.03)$ & $-0.048(0.03)$ & $-0.049(0.02)$ & $-0.062(0.04)$ & $-0.067(0.03)$ & $-0.068(0.03)$ \\ \hline
      GMD2 & $0.276(0.27)$ & $0.353(0.24)$ & $0.426(0.22)$ & $0.342(0.28)$ & $0.428(0.25)$ & $0.508(0.21)$ \\ \hline 
      GMD3 &  $-$ & $-$ & $0.403 (0.27)$ & $-$ & $-$ & $0.464 (0.28)$ \\ \hline
      GMDM &  $-$ & $-$ & $0.435 (0.24)$ & $-$ & $-$ & $0.522 (0.24)$ \\ \hline
    \end{tabular}}
    \caption{Mean(SD) MCC values for experiment 2 using $\alpha = 0.05$.}
    \label{tab:exp2_mcc05}
\end{table}

\begin{table}[htbp]
    \centering
    \Large
    \resizebox{\textwidth}{!}{\begin{tabular}{|c|c|c|c|c|c|c|} \hline
      & \multicolumn{3}{|c|}{$5\%$ Outliers in Test Data}   & \multicolumn{3}{|c|}{$10\%$ Outliers in Test Data} \\ \hline
      Method & $n=50$ & $n=100$ & $n=250$ & $n=50$ & $n=100$ & $n=250$ \\ \hline
      EFDM & $0\textbf{.601(0.16)}$ & $\textbf{0.589(0.14)}$ & $\textbf{0.578(0.10)}$ & $\textbf{0.710(0.14)}$ & $\textbf{0.702(0.11)}$ & $\textbf{0.697(0.09)}$ \\ \hline
      SNCM1 & $-0.046(0.07)$ & $-0.051(0.06)$ & $-0.059(0.05)$ & $-0.066(0.08)$ & $-0.081(0.06)$ & $-0.085(0.05)$ \\ \hline
      SNCM2 & $-0.042(0.07)$ & $-0.048(0.07)$ & $-0.059(0.05)$ & $-0.059(0.08)$ & $-0.070(0.06)$ & $-0.085(0.05)$ \\ \hline
      GMD1 & $-0.070(0.03)$ & $-0.073(0.02)$ & $-0.072(0.02)$ & $-0.098(0.04)$ & $-0.103(0.04)$ & $-0.102(0.03)$ \\ \hline
      GMD2 & $0.359(0.23)$ & $0.444(0.17)$ & $0.495(0.12)$ & $0.440(0.26)$ & $0.546(0.17)$ & $0.609(0.13)$ \\ \hline 
      GMD3 &  $-$ & $-$ & $0.479 (0.16)$ & $-$ & $-$ & $0.588 (0.15)$ \\ \hline
      GMDM &  $-$ & $-$ & $0.473 (0.11)$ & $-$ & $-$ & $0.596 (0.11)$ \\ \hline
    \end{tabular}}
    \caption{Mean(SD) MCC values for experiment 2 using $\alpha = 0.10$.}
    \label{tab:exp2_mcc10}
\end{table}

Both Tables \ref{tab:exp2_mcc05} and \ref{tab:exp2_mcc10} exhibit similar patterns to those seen in Table \ref{tab:exp1_nrdp_covg}. As in the previous experiment, EFDM outperforms all other methods in each case. The performance gap between EFDM and the other methods is larger when $\alpha = 0.05$ than when $\alpha=0.10$. In fact, when $\alpha=0.10$ the mean MCC for EFDM is within one standard deviation of the mean MCC for GMD2 (and GMD3 and GMDM when applicable). In both tables, SNCM1, SNCM2, and GMD1 consistently perform poorly, registering negative MCC values in every case.

\subsection{Experiment 3} \label{sim:exp3}
In the final experiment, we aim to simulate the second outlier detection scenario, where only a single data set is available and the user desires to identify the most outlying observations. Here we randomly select 100 of the test data sets from experiment 2 where the outlier rate is $5\%$ and 100 where the outlier rate is $10\%$. For each of these data sets, we use the leave-one-out versions, described in Section \ref{efdm:loo}, of the methods in Table \ref{tab:exp3_methods}. Note that the parameters for the GMD methods in this experiment were selected to provide greater flexibility than those used in experiments 1 and 2 while still being able to fit the GMM parameters. 

\begin{table}[htbp]
    \centering
    \resizebox{\textwidth}{!}{
    \begin{tabular}{r|l} 
        Method Name & Details \\ \hline
        EFDM & NCM uses amplitude and phase distances only \\
        SNCM1 & Uses the cross-sectional mean function and the optimal modulation function \\
        SNCM2 & Uses the Karcher mean and the optimal modulation function \\
        GMD1 & Uses $p = 10$, $K = 1$ \\
        GMDM1 & Uses $p = 10$, $K = 1$ \\
        GMD2 & Uses $p = 5$, $K = 2$ \\
        GMDM2 & Uses $p=5$, $K = 2$ \\
        GMD3 & Uses $p = 4$, $K = 3$ \\
        GMDM3 & Uses $p=4$, $K = 3$ \\
    \end{tabular}}
    \caption{Description of methods used for experiment 3.}
    \label{tab:exp3_methods}
\end{table}

The leave-one-out approach produces a label for each observation in a given data set. We again evaluate method performance using MCC. Mean MCC and standard deviation over the 100 data sets (per outlier rate) are given in Table \ref{tab:exp3_mcc} As in experiment 2, we used both $\alpha = 0.05$ and $\alpha = 0.10$ as levels of significance. Additional metrics from this experiment are given in Appendix \ref{appdx}.

\begin{table}[htbp]
    \centering
    \begin{tabular}{|c|c|c|c|c|} \hline
      &  \multicolumn{2}{|c|}{$\alpha = 0.05$}  & \multicolumn{2}{|c|}{$\alpha = 0.10$} \\ \hline
    Method & $5\%$ Outlier Rate & $10\%$ Outlier Rate & $5\%$ Outlier Rate & $10\%$ Outlier Rate \\ \hline
    EFDM & $\textbf{0.744(0.13)}$ & $\textbf{0.630(0.11)}$ & $\textbf{0.678(0.07)}$ & $\textbf{0.800(0.08)}$ \\ \hline
    SNCM1 & $-0.051(0.01)$ & $-0.071 (0.02)$ & $-0.066(0.04)$ & $-0.101(0.03)$ \\ \hline
    SNCM2 & $-0.048(0.03)$ & $-0.064 (0.04)$ & $-0.056(0.5)$ & $-0.094(0.05)$ \\ \hline
    GMD1 & $0.024(0.11)$ & $-0.067(0.04)$ & $0.092(0.12)$ & $-0.058(0.07)$ \\ \hline
    GMDM1 & $0.021(0.10)$ & $-0.066(0.04)$ & $0.088(0.12)$ & $-0.058(0.07)$ \\ \hline
    GMD2 & $0.206(0.18)$ & $0.023(0.11)$ & $0.310(0.14)$ & $0.047(0.11)$ \\ \hline
    GMDM2 & $0.200(0.17)$ & $0.021(0.11)$ & $0.299(0.14)$ & $0.055(0.11)$ \\ \hline
    GMD3 & $0.102(0.16)$ & $-0.041(0.07)$ & $0.160(0.14)$ & $-0.053(0.08)$ \\ \hline
    GMDM3 & $0.103(0.15)$ & $-0.046(0.07)$ & $0.157(0.15)$ & $-0.050(0.082)$ \\ \hline
    \end{tabular}
    \caption{Mean(SD) MCC values for experiment 3.}
    \label{tab:exp3_mcc}
\end{table}

In this case, EFDM far outperforms every other method in each case. For all GMD/GMDM methods, performance is always better with the $5\%$ outlier rate than with the $10\%$ outlier rate. EFDM, on the other hand, performs better when $\alpha$ matches the outlier rate. The SNCM methods again consistently perform poorly, and the GMD/GMDM methods perform either similarly or only slightly better.

\subsection{Experiment Conclusions} \label{sim:exp_conc}
There are several conclusions to be drawn from these results. Before doing so, we first emphasize that these experiments are at least somewhat removed from the original settings in which the methods of \cite{lei2015} and \cite{diquigiovanni2021} were introduced. As discussed, shape outliers are harder to detect than magnitude outliers, and we intended the narrow and double peaked outliers to be a particularly difficult case of shape outliers relative to our standard simulated data. When plotted, both narrow and double peaked outliers fit well within the standard data. The narrow data are also only very slightly different in shape than the standard data. Thus, we do not believe that our results can be used to make statements about the performance of these methods in all possible use cases.

Regarding SNCM, there does not seem to be any benefit of using the Karcher mean rather than a cross-sectional mean function as SNCM1 and SNCM2 performed similarly in all experiments. It is also not well suited for detecting the kind of shape outliers that we used in these experiments. 

There did not seem to be much of a difference in performance between GMD and GMDM methods when they used the same parameters. In experiments 1 and 2 (corresponding to the first outlier detection scenario), some of the GMD/GMDM methods perform reasonably well detecting these difficult shape outliers. We believe that the poor performance of GMD1 in experiments 1 and 2 is due to its rigidity. That is, the projected functional data are being represented by a single bivariate Gaussian distribution. It is likely not flexible enough to provide a good representation of the functional data we used. As flexibility increases with larger values of $p$ and $K$, these methods do much better, even coming fairly close to EFDM performance on experiment 2. With even greater flexibility, it is possible that GMD/GMDM methods could perform even better. One downside of this method, as seen in experiments 1 and 2, is that more data are needed to enable more flexible implementations. On the other hand, GMD/GMDM methods are trained and produce labels with relatively low computational expense, so they may be a good choice with large functional data sets. For the second outlier detection scenario explored in experiment 3, however, the results are not encouraging. We hypothesize that GMD/GMDM performs poorly in the second outlier detection scenario due to the projection being achieved with FPCA. Specifically, that the eigendecomposition we utilized to find the functional principal component is more susceptible to the effects of outlier contamination. We believe that further investigation into the leave-one-out versions of GMD/GMDM methods is needed before it can be recommended for use.

We have already noted that EFDM performs well in each experiment. Because the leave-one-out approach loses the coverage guarantee that comes with having exchangeable training/calibration data, we were somewhat surprised by how well EFDM performed in experiment 3. These results suggest that EFDM can be useful for both outlier detection scenarios. We also emphasize that EFDM can perform well on relatively small sample sizes. When $n=50$, $n_1 = 34$ and $n_2 = 16$. Providing the correct coverage and being useful as an outlier detector with such a small amount of data is an important feature of EFDM.

Finally, experiments 2 and 3 clearly demonstrate that results depend heavily on the selection of the significance level, $\alpha$. In these methods, $\alpha$ has the traditional frequentist interpretation of the false positive rate. Thus, to select an appropriate value in a real application, the user needs to understand and compare the costs of false positives and false negatives. This decision is further influenced by the fact that we must choose $\alpha \ge \frac{1}{n_2+1}$. Because selecting $\alpha$ forces a determination about outlier status, it may be wise to analyze p-values rather than settling on a single value of $\alpha$ for a given application. 

\section{Analysis of Exemplar Data} \label{exem}
In this section, we analyze two real data sets using EFDM. For both exemplars, we EFDM is applied in both outlier detection scenarios. The purpose of these exemplars is to demonstrate use cases of EFDM and not to provide thorough analyses of the data sets.

\subsection{Zener Diode Data} \label{exem:diode}
\begin{figure}[htbp]
    \centering
    \includegraphics[width=0.75\linewidth]{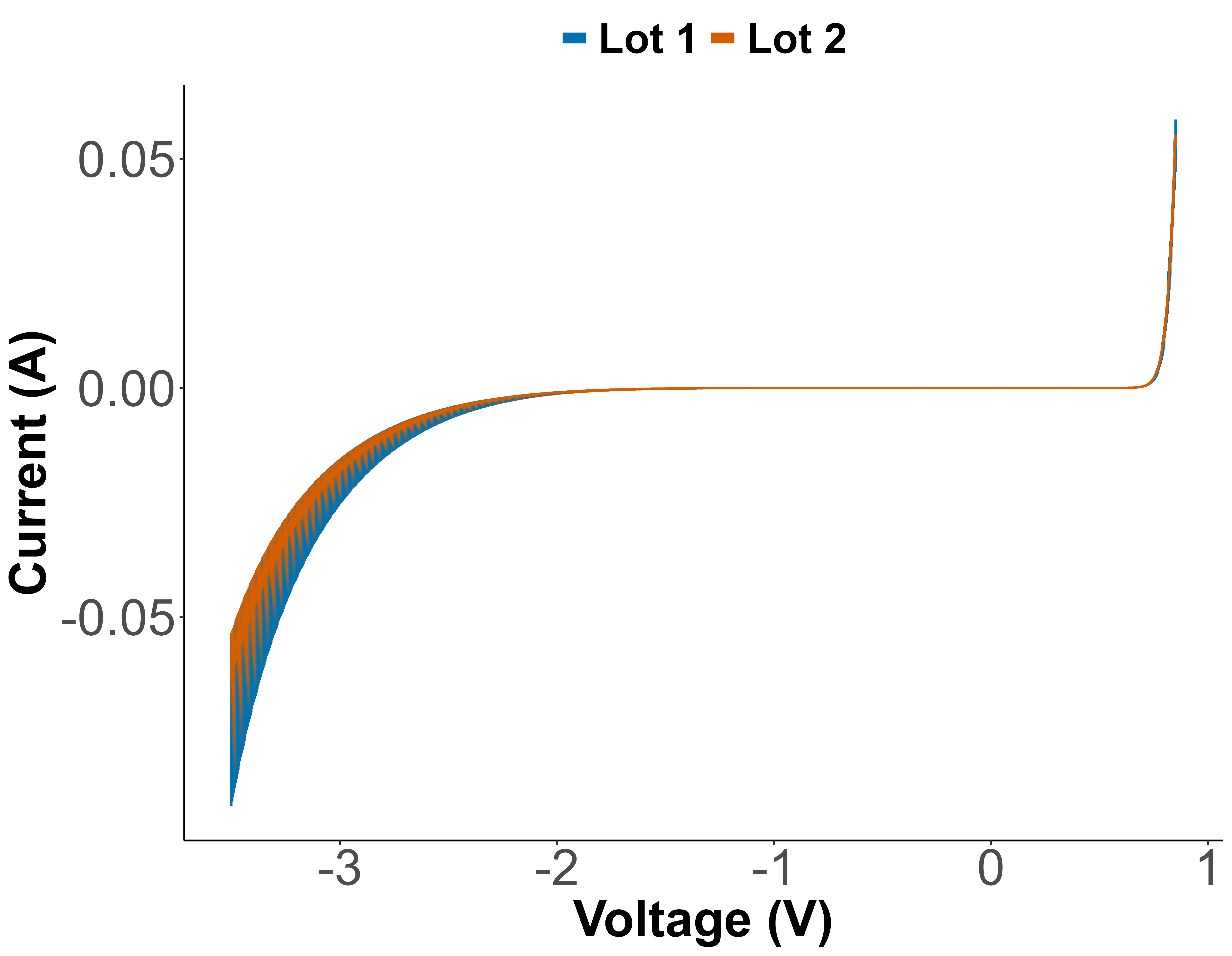}
    \caption{I-V sweeps from the 294 Zener diodes where the color represents the production lot for each device.}
    \label{fig:diodes}
\end{figure}

In the work of \cite{champon2023}, functional data methods were used to analyze data from 193 MMSZ522BT1G Zener diodes. The diodes were produced in two separate production lots, and three different methods were used to analyze the data. The data are in the form of current-voltage (I-V) sweeps which are produced by forcing a voltage and measuring the current. As seen in Figure \ref{fig:diodes}, it is natural to treat these curves as functional data. Because none of these curves appear to be simply vertically shifted magnitude outliers, we do not include the translation distance in the NCM for this analysis.

 In this work, we use a superset of the data from \cite{champon2023} to further investigate two of the questions from their previous work. What we call Lot 1 produced 145 of these devices while the remaining 149 came from Lot 2. The first question we consider is if we can distinguish between the lots. Ideally, device behavior would be identical between the lots, but \cite{champon2023} used FPCA to show that the data are easily separable on a simple two-dimensional projection space. The second question we consider is if there are outliers within the production lots. In the previous work, pairwise amplitude and phase distances between all I-V sweeps were computed. These distances were used to look for differences both within and between the production lots. The results suggested there may be some outlying devices within Lot 1. The first question can be addressed using standard EFDM while the second can be addressed with the leave-one-out version.

\begin{figure}[htbp]
    \centering
    \begin{tabular}{cc}
      \includegraphics[width=0.5\linewidth]{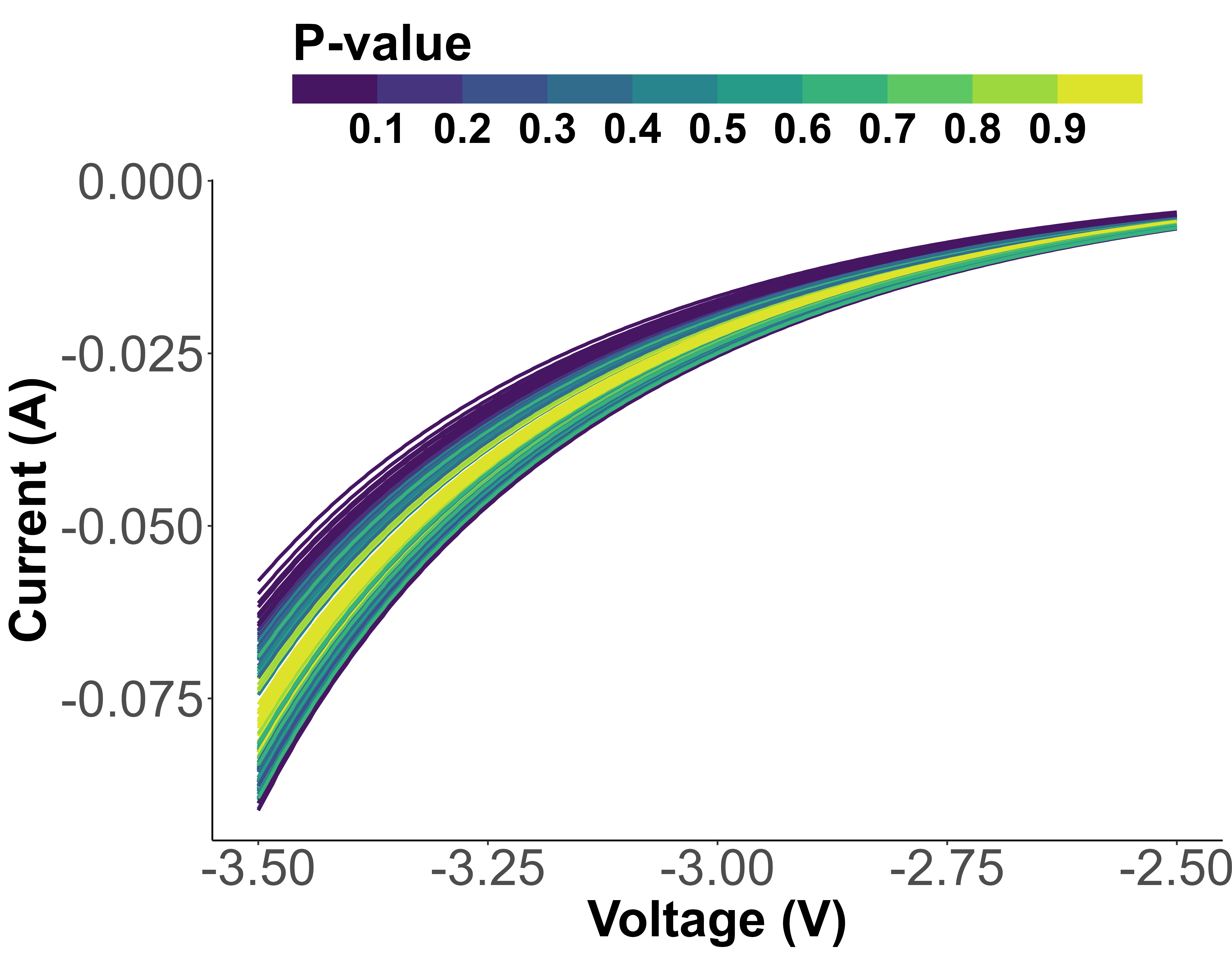}   &  \includegraphics[width=0.5\linewidth]{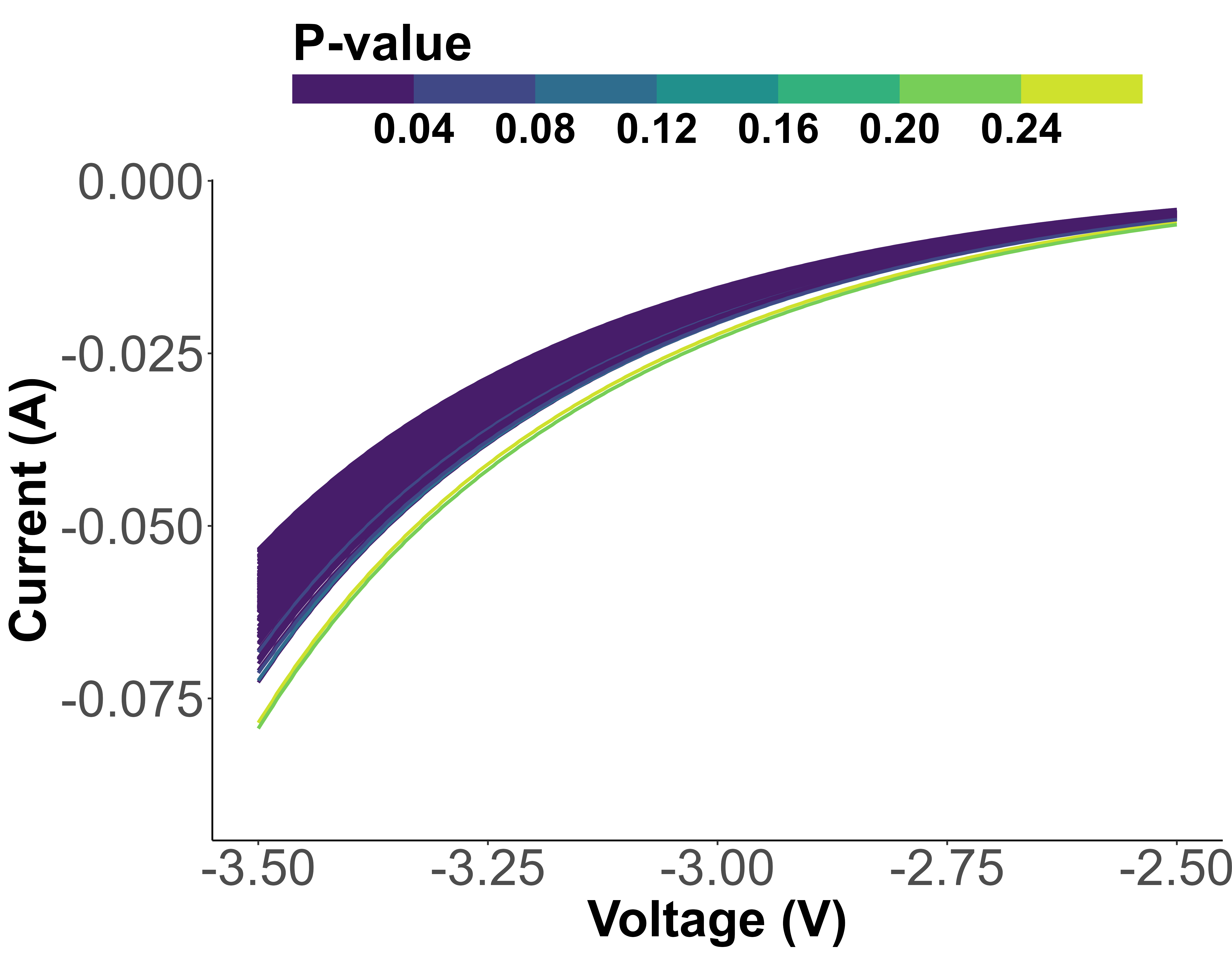} \\
      (a) & (b) \\
      \includegraphics[width=0.5\linewidth]{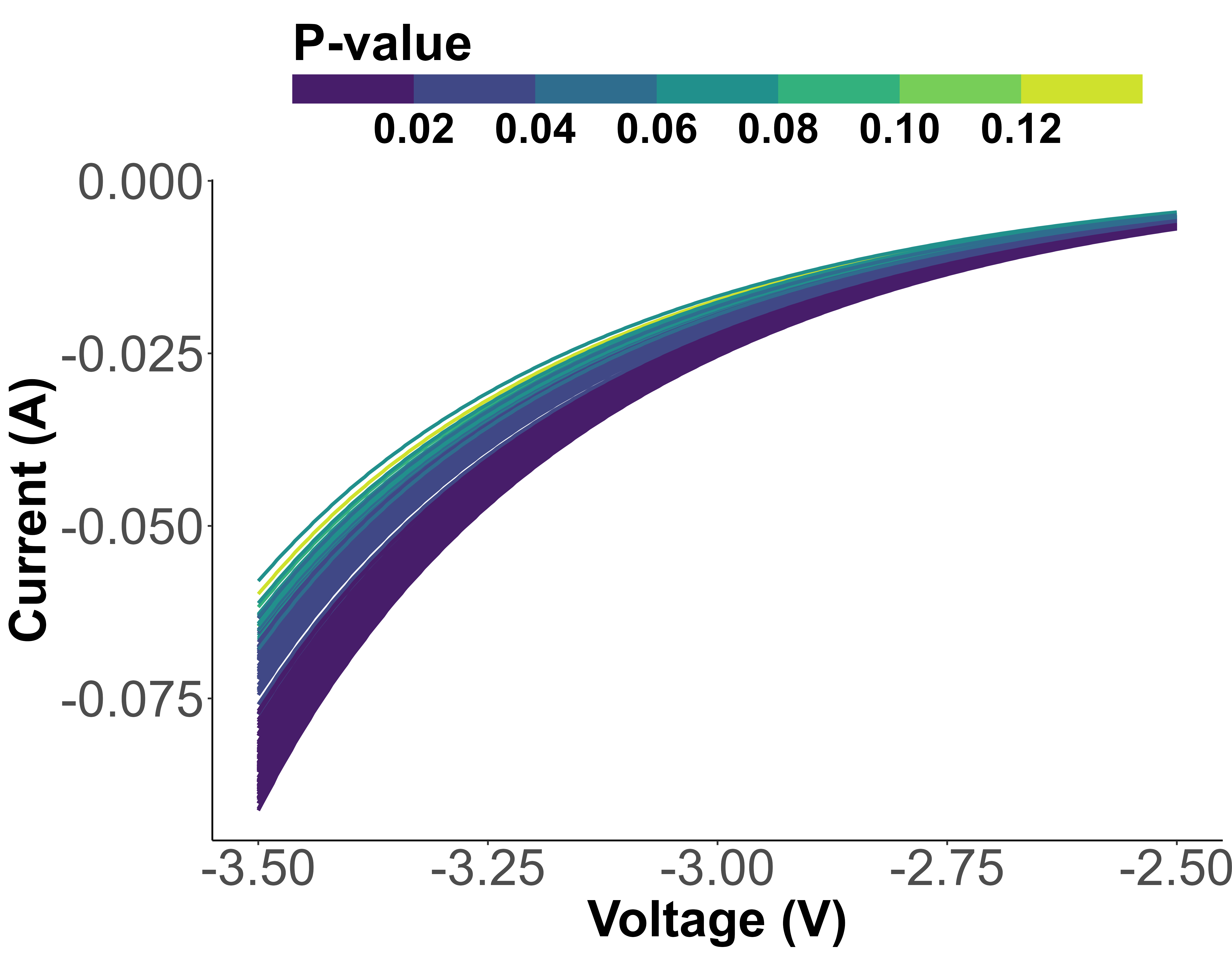}   &  \includegraphics[width=0.5\linewidth]{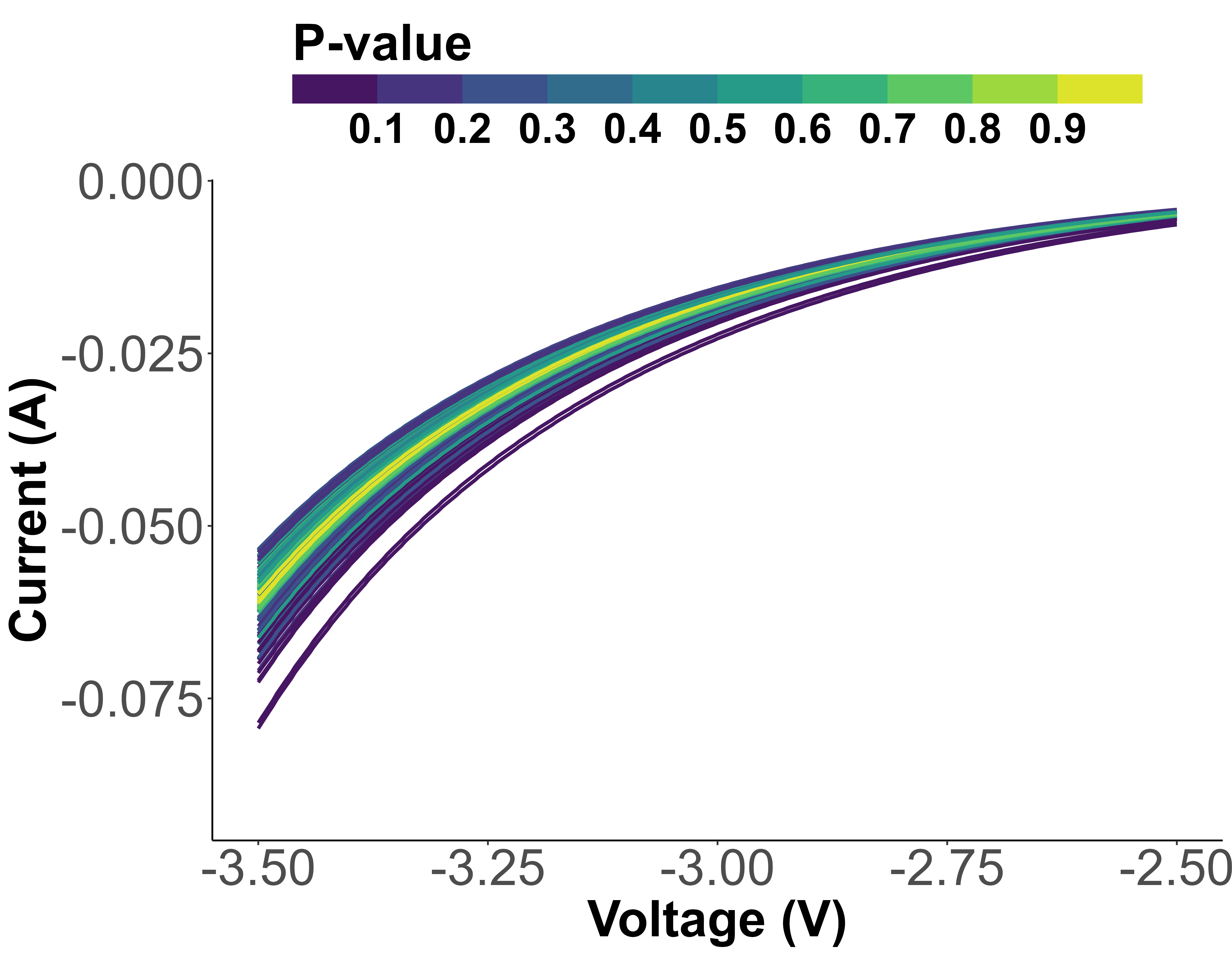} \\
      (c) & (d)
    \end{tabular}
    \caption{I-V sweeps where the color corresponds to the p-value produced by EFDM. (a) Lot 1 p-values when training on Lot 1 data; (b) Lot 2 p-values when training on Lot 1 data; (c) Lot 1 p-values when training on Lot 2 data; (d) Lot 2 p-values when training on Lot 2 data.}
    \label{fig:diode_pval}
\end{figure}

In our analysis, to address the first question, EFDM is first trained on Lot 1 data and then used to label Lot 2 data as either inliers or outliers. This is also then done using Lot 2 as training data and labeling Lot 1 data. For the second question, we use our leave-one-out approach in each lot separately to identify the most outlying devices within each lot. Results from these analyses are given in Table \ref{tab:diode_outliers} and Figure \ref{fig:diode_pval}. As is clearly seen in Figure \ref{fig:diodes}, most of the variability between functions occurs between the minimum voltage (-3.5V) and roughly -2.5V (this was confirmed by the FPCA of \citealp{champon2023}). For this reason, and to make the plots more informative, we only plot this region (the entire functions are used within EFDM to produce these results). 

In Figure \ref{fig:diode_pval}, the colors of the plotted functions correspond to the p-values produced by EFDM. In the first row of Figure \ref{fig:diode_pval}, panels (a) and (b), the data from Lot 1 is used for training. The difference between panels (a) and (b) is that the testing data in panel (b) is the data from Lot 2 while in panel (a) the leave-one-out approach is used. Panels (c) and (d) in Figure \ref{fig:diode_pval} use Lot 2 for training and the testing data is Lot 1 in panel (c) and Lot 2 in panel (d). Panels (a) and (d) contain the results from the leave-one-out procedure while panels (b) and (c) contain results from the standard EFDM. Panels (a) and (d) match our intuition that the most outlying functions (smaller p-values) will be those furthest from the center of the data. In panels (b) and (c), which display Lot 2 and Lot 1 data, respectively, the only functions with higher p-values are those at the bottom in panel (b) and at the top in panel (c). This is where the two lots overlap (see Figure \ref{fig:diodes}), so it makes sense that the only Lot 2 functions that could reasonably be inliers for Lot 1 are those that overlap with Lot 1 functions (and vice versa). Note, however, that the p-value color scales differ between the panels. The scales for (a) and (d) are similar to each. While the ranges for (b) and (c) are both much smaller than (a) and (d), the range of (c) is only about half the range for (b). While the method producing the diagonal plots (leave-one-out) is different than the method that produces the off-diagonal plots, this seems to suggest that devices within a given lot are much more similar than devices across the lots.

\begin{table}[htbp]
    \centering
    \begin{tabular}{|c|c|c|c|c|} \hline
     & \multicolumn{2}{|c|}{$\alpha = 0.05$}    &  \multicolumn{2}{|c|}{$\alpha = 0.10$} \\ \hline
      & Test on Lot 1 & Test on Lot 2 & Test on Lot 1 & Test on Lot 2 \\ \hline
     Train on Lot 1 & $0.028$ & $0.980$ & $0.103$ & $0.987$ \\ \hline 
     Train on Lot 2 & $0.862$ & $0.040$ & $0.986$ & $0.107$ \\ \hline
    \end{tabular}
    \caption{Percentages of outliers in the testing set when using the designated lots for training and testing with two different levels of significance, $\alpha = 0.05$ and $\alpha = 0.10$.}
    \label{tab:diode_outliers}
\end{table}

Table \ref{tab:diode_outliers} gives the percentage of functions labeled as outliers when training on one lot for two different levels of significance, $\alpha = 0.05$ and $\alpha = 0.10$ (since the smaller lot contains 145 diodes, the calibration set size is 48 and the minimum $\alpha$ that can be used is $\frac{1}{49} \approx 0.02$; we simply select the commonly used values of $0.05$ and $0.10$ that are allowed with this minimum). With either level of significance, Lot 2 functions are almost always labeled as outliers when training on Lot 1 functions. When training on Lot 2 and testing on Lot 1, there is a larger difference in the percentage of outliers between the two significance levels, but both values are still fairly high. Thus, this seems to further support the conclusion of \cite{champon2023} that the device behavior, as contained in the I-V sweeps, allows us to distinguish fairly well between the two lots. 

For the second analysis question, the two significance levels give quite different values for the within-lot outlier percentages. When $\alpha = 0.10$, roughly $10\%$ of each lot are labeled as outliers. When using $\alpha = 0.05$, only about $3\%$ of Lot 1 and $4\%$ of Lot 2 are labeled as outliers. Without additional information about the costs of false positives and false negatives, it is perhaps unwise to select a single value of $\alpha$. The corresponding plots of Figure \ref{fig:diode_pval} are likely more useful here as they show p-values without forcing a decision to be made. In any case, the number of within-lot outliers in Lot 1 does not show up to the same degree as found by \cite{champon2023}. Whether this is because we use more data (perhaps most or all of the new Lot 1 data we use are inliers, thus driving down the overall outlier percentage) or because the methods are too different (the previous work used pairwise distances instead of our distance to lot Karcher means), more investigation would be needed before definite conclusions can be made.

\subsection{Temperature Data} \label{exem:temp}
Our second exemplar uses data downloaded from the National Centers for Environmental Information at the National Oceanic and Atmospheric Administration (NOAA). We downloaded daily high and low temperatures and calculated the daily mean temperature from the years 1981-2024 at five different airports in the United States: Atlanta, Dallas-Love Field, Dallas-Fort Worth, Miami, and San Francisco. In this case, each year is taken as a single functional observation. As with the previous exemplar, our goal is simply to demonstrate EFDM being used in both outlier detection scenarios. So, as before, we train on the data from each site and then label the years at all other sites as inliers or outliers. With this analysis, we hope to see most years labeled as inliers for similar sites (e.g., the two sites in Dallas) and most years labeled as outliers when the sites are dissimilar (e.g., Miami and San Francisco). We also consider the second outlier detection scenario by using the leave-one-out approach to predict the most outlying years at each of the sites. Because the data from San Francisco and Miami look very similar in shape but different in magnitude, we include the translation distance as described in Section \ref{efdm} in the NCM.

\begin{figure}[htbp]
    \centering
    \begin{tabular}{c}
      \includegraphics[width = \linewidth]{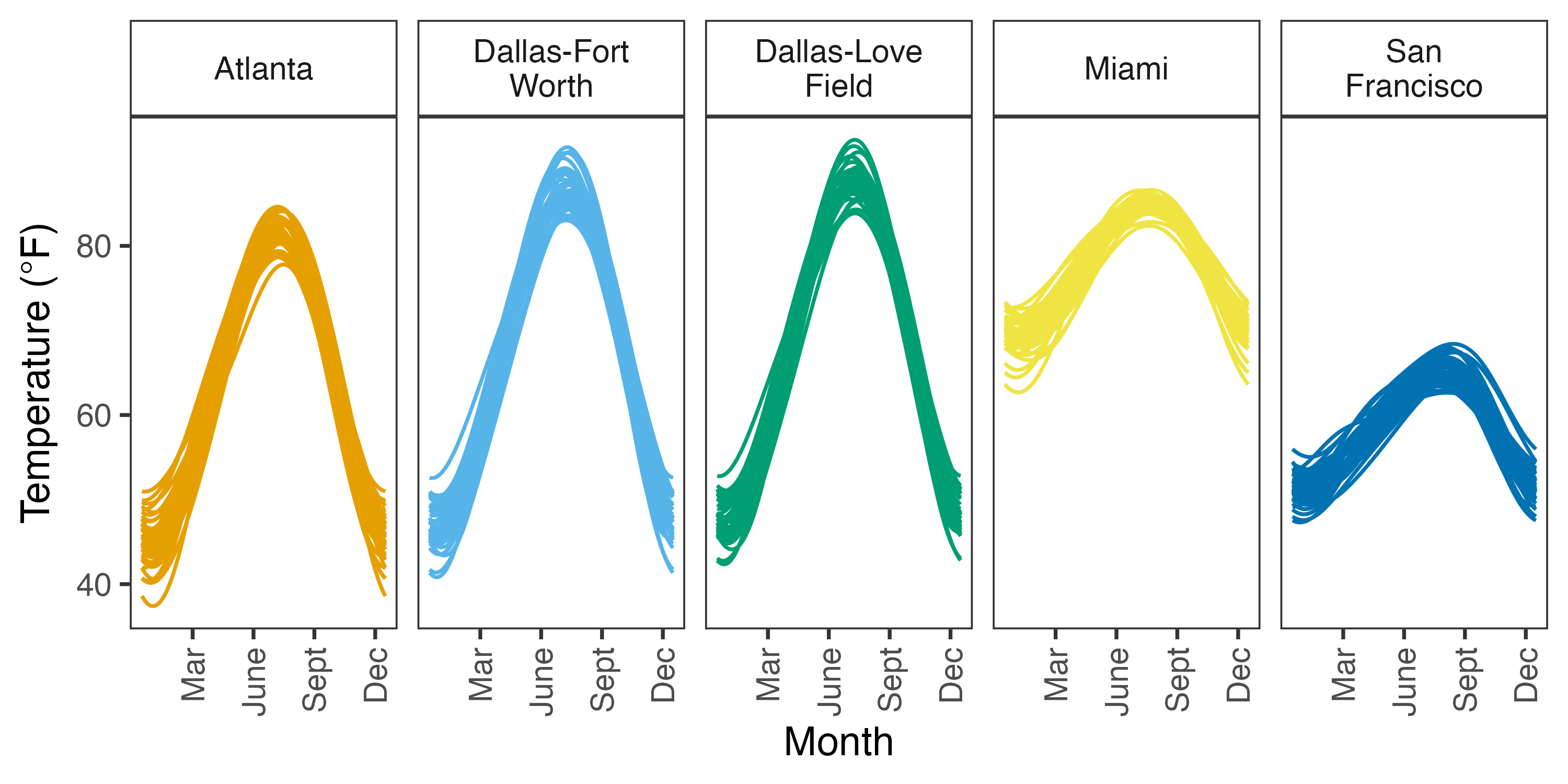} \\
      (a) \\
      \includegraphics[width = \linewidth]{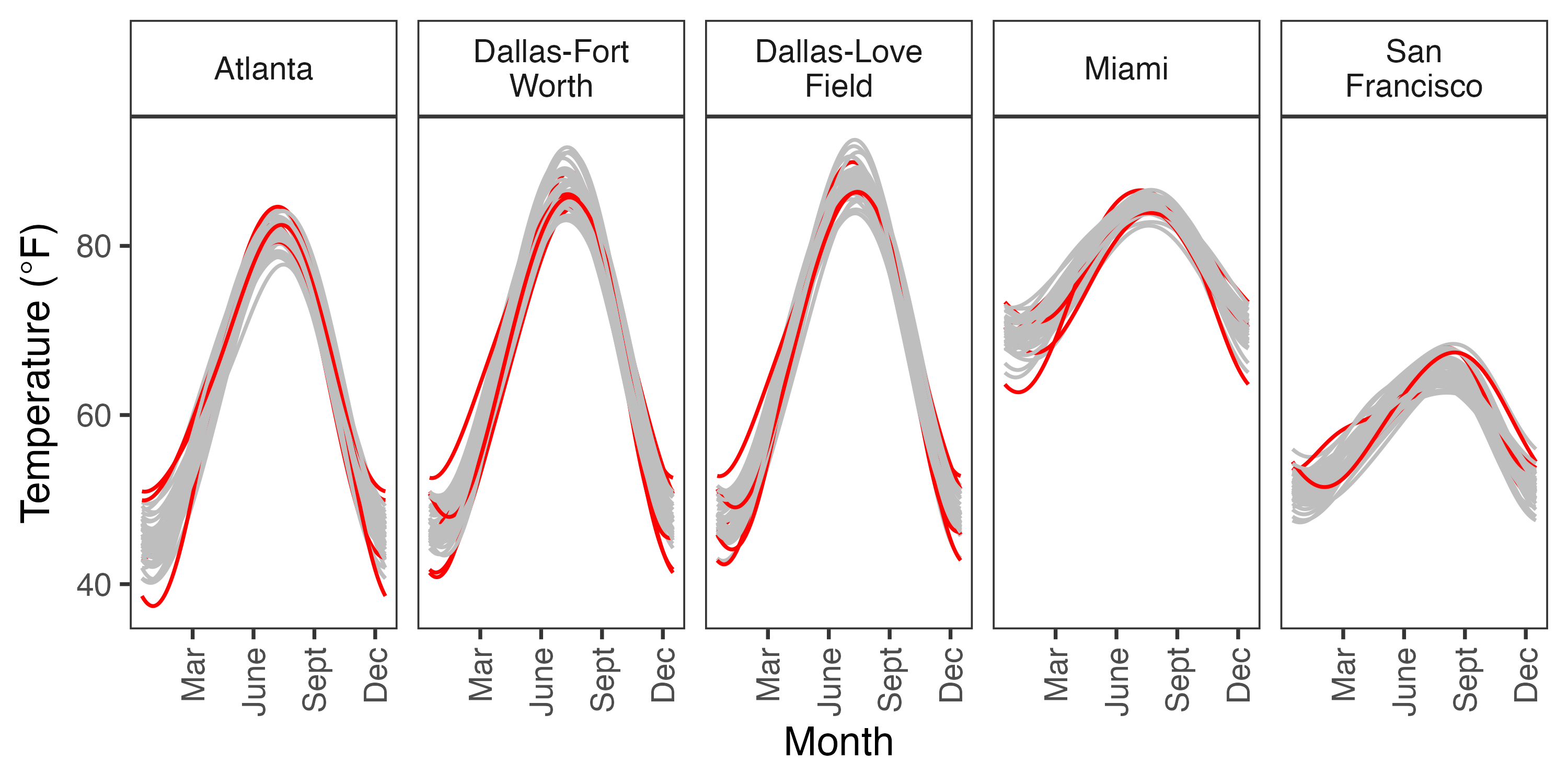}    \\
      (b)
    \end{tabular}
    
    \caption{(a) The monthly averaged daily mean temperature data for five sites in the United States. (b) The same data where the red curves are those labeled as outliers using the leave-one-out approach within each site.}
    \label{fig:temp_dat}
\end{figure}

EFDA methods work best on smooth data, and since our goal is to demonstrate EFDM capabilities on real data rather than provide thorough analysis of the exemplar data, the temperature data went through overly simplistic preprocessing before training. We stress that the analysis here should not be used to make general conclusions about the temperature data themselves but only to shed light on EFDM as a method. To produce the data seen in Figure \ref{fig:temp_dat}(a), we first computed a daily mean temperature, $d_{mean} = 0.5(d_{max} + d_{min})$ for every day over the 44 years at each site, where $d_{max}$ and $d_{min}$ are the daily high and daily low temperatures, respectively. We then averaged all $d_{mean}$ temperatures in a given month to produce $m_{mean}$. At this point, each functional observation (year) was observed over 12 time points (months). We then smoothed the data using a Fourier basis, as described by \cite{Ramsay2005}, and interpolated the smoothed functions so that we again had daily mean temperatures. Thus the data in Figure \ref{fig:temp_dat}(a) are daily mean temperatures produced from monthly averages.

Because each site has $n=44$ years of temperature data, we have $n_1 = 30$ training functions and $n_2 = 14$ calibration functions. Since we need $\alpha \ge \frac{1}{n_2+1}$, we use $\alpha = 0.10$. Table \ref{tab:temp_res} shows the percentage of years at the site in the column that are labeled as outliers when training on the site in the row. In this table, the off-diagonals correspond to the standard EFDM ICAD approach while the diagonal elements come from our leave-one-out approach. As we might expect from Figure \ref{fig:temp_dat}, all years at all other sites are labeled as outliers when training on either Miami or San Francisco. When training and testing on Atlanta and the two Dallas sites, the proportion of outliers is much smaller in each case. These results match our intuition about which sites should be similar to each other.

\begin{table}[htbp]
    \centering
    \begin{tabular}{|c|c|c|c|c|c|c|} \hline
      & \multicolumn{5}{|c|}{Testing Site} \\ \hline
     Training Site & ATL & DFW & DAL & MIA & SFO \\ \hline
     ATL  & $0.091$ & $0.068$ & $0.068$ & $1$ & $1$ \\ \hline
     DFW  & $0.25$ & $0.136$ & $0.114$ & $1$ & $1$ \\ \hline
     DAL  & $0.136$ & $0.114$ & $0.114$ & $1$ & $1$ \\ \hline
     MIA  & $1$ & $1$ & $1$ & $0.114$ & $1$ \\ \hline
     SFO  & $1$ & $1$ & $1$ & $1$ & $0.091$ \\ \hline
    \end{tabular}
    \caption{Percentage of outliers when training on the site in the row and labeling the site in the column for all five sites: Atlanta (ATL), Dallas-Forth Worth (DFW), Dallas-Love Field (DAL), Miami (MIA), and San Francisco (SFO).}
    \label{tab:temp_res}
\end{table}

The results of the leave-one-out approach to identifying outliers within a site are shown in both the diagonal of Table \ref{tab:temp_res} and Figure \ref{fig:temp_dat}(b) where the red functions are the ones labeled as outliers. From a visual inspection, the red functions appear to be either on the edges (in terms of either amplitude or phase) or they look to have markedly different shapes than some of the other functions. This is perhaps clearest in the Miami site where two of the red functions seem to rise more steeply from about March to June than the rest of the functions. Thus, in both outlier detection scenarios, the results appear to match intuition and visual inspection of the data.

\section{Conclusion} \label{conclusion}
In this paper, we have introduced the elastic functional distance metrics ICAD method, evaluated it alongside two competing ICAD methods for functional data, and demonstrated its effectiveness on two real data sets. We argue that our results show EFDM to be effective at detecting shape and magnitude outliers in both outlier detection scenarios. 

As argued by \cite{bates2023}, users of conformal prediction methods, including ICAD, need to be aware of potential multiple testing pitfalls. In particular, the coverage guarantee of CP methods holds for a single test point, $P(Y_{n+1} \in C(X_{n+1})) \ge 1-\alpha$. However, the probability that a large number of test points will all belong to their respective prediction sets can become much smaller than $1-\alpha$ as the number of test points increases. This is the same problem encountered with the family-wise error rate in standard hypothesis testing \citep{shaffer1995}. We did not implement any corrections for this (such as those discussed by \citealp{bates2023}), and this could be potentially increasing the number of false positives for some of our results. In future work, we will implement such corrections and compare the outcomes with our current results.

\section*{Code and Data}
The R code for implementing the methods and analyses in this work is available at \url{https://github.com/sandialabs/conformal-functional-outliers}. This repository also contains data sets used in both the exemplars and the simulation experiments.

\acks{The authors express gratitude to Tom Buchheit and Dave Canfield for producing and sharing the Zener diode data. We also wish to thank Dr. Jing Lei who graciously shared \texttt{R} code implementing his functional conformal prediction methods.

This work was supported by the Laboratory Directed Research and Development program at Sandia
National Laboratories, a multimission laboratory managed and operated by National Technology and
Engineering Solutions of Sandia LLC, a wholly owned subsidiary of Honeywell International Inc. for the U.S.
Department of Energy’s National Nuclear Security Administration under contract DE-NA0003525. This paper describes objective technical results and analysis. Any subjective views or opinions that might be expressed in the paper do not necessarily represent the views of the U.S. Department of Energy or the United States Government.

This article has been authored by an employee of National Technology \& Engineering Solutions of Sandia, LLC under Contract No. DE-NA0003525 with the U.S. Department of Energy (DOE). The employee owns all right, title and interest in and to the article and is solely responsible for its contents. The United States Government retains and the publisher, by accepting the article for publication, acknowledges that the United States Government retains a non-exclusive, paid-up, irrevocable, world-wide license to publish or reproduce the published form of this article or allow others to do so, for United States Government purposes. The DOE will provide public access to these results of federally sponsored research in accordance with the DOE Public Access Plan https://www.energy.gov/downloads/doe-public-access-plan .

}

\bibliography{references}

\appendix
\section{Additional Experiment Results} \label{appdx}
Here we provide additional results from each experiment in sections \ref{sim:exp1} through \ref{sim:exp3}.

\subsection{Experiment 1 Coverage Plots}
\begin{figure}[htbp]
    \centering
    \begin{tabular}{cc}
      \includegraphics[width=0.50\linewidth]{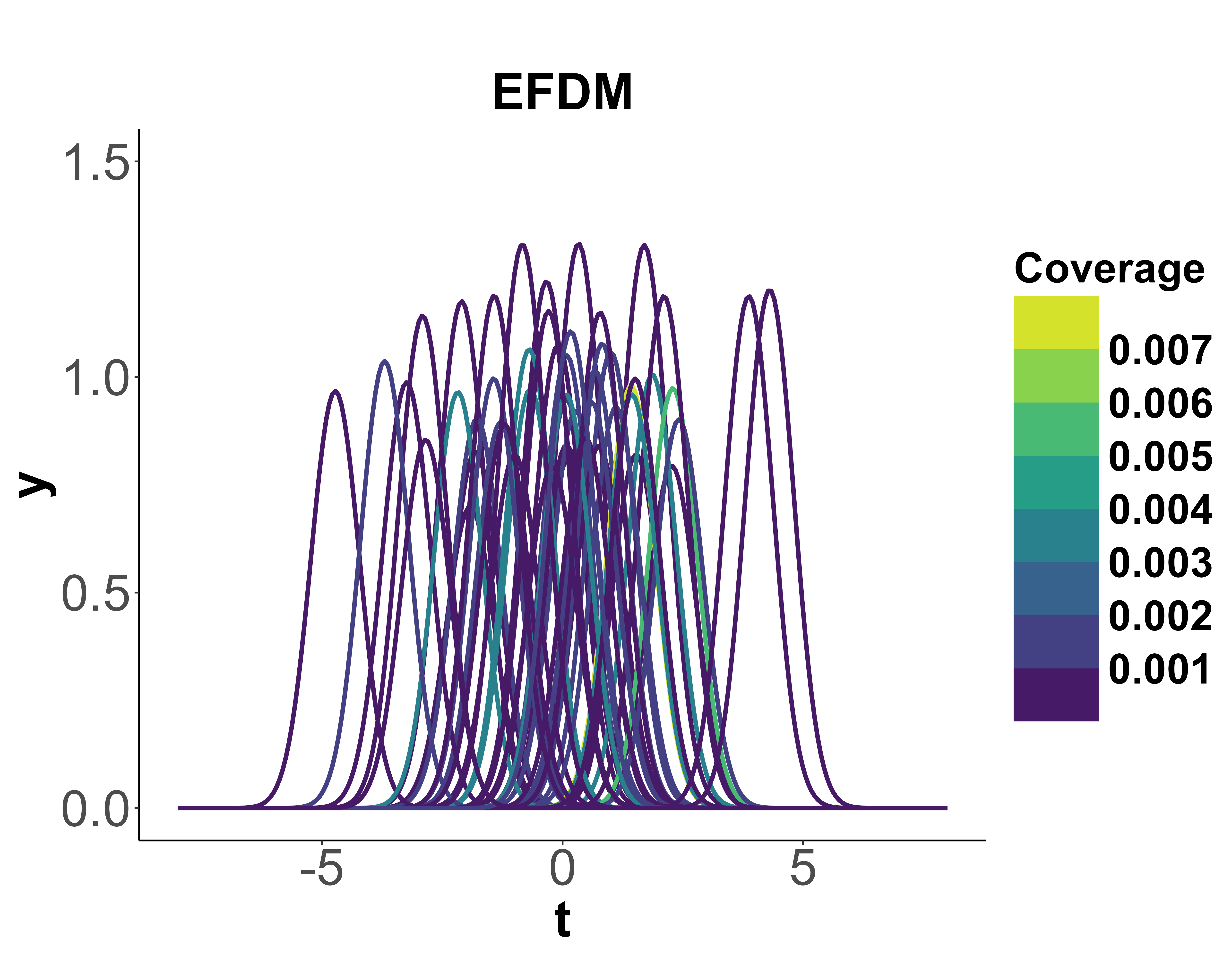} &  
      \includegraphics[width=0.50\linewidth]{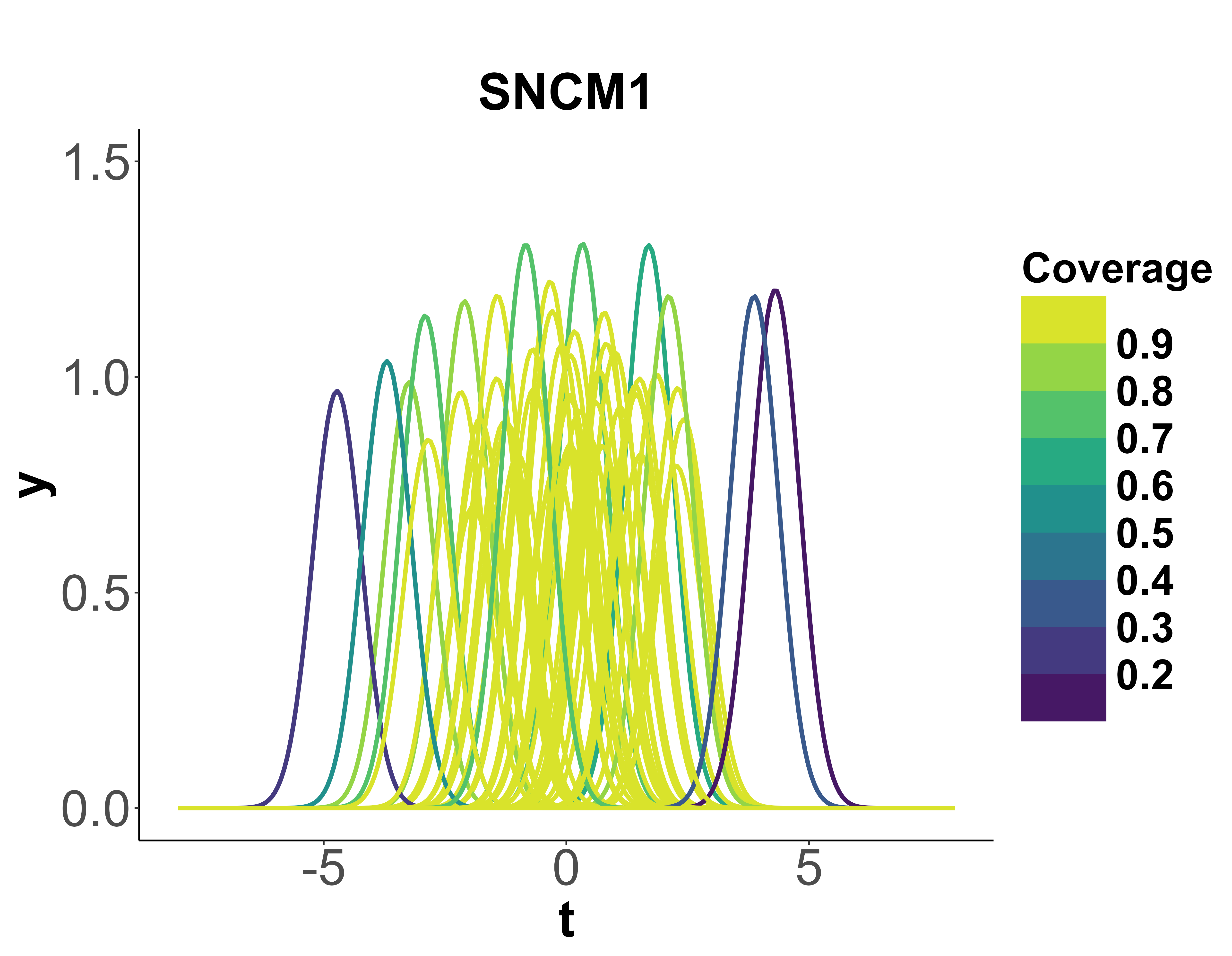} \\
      \includegraphics[width=0.50\linewidth]{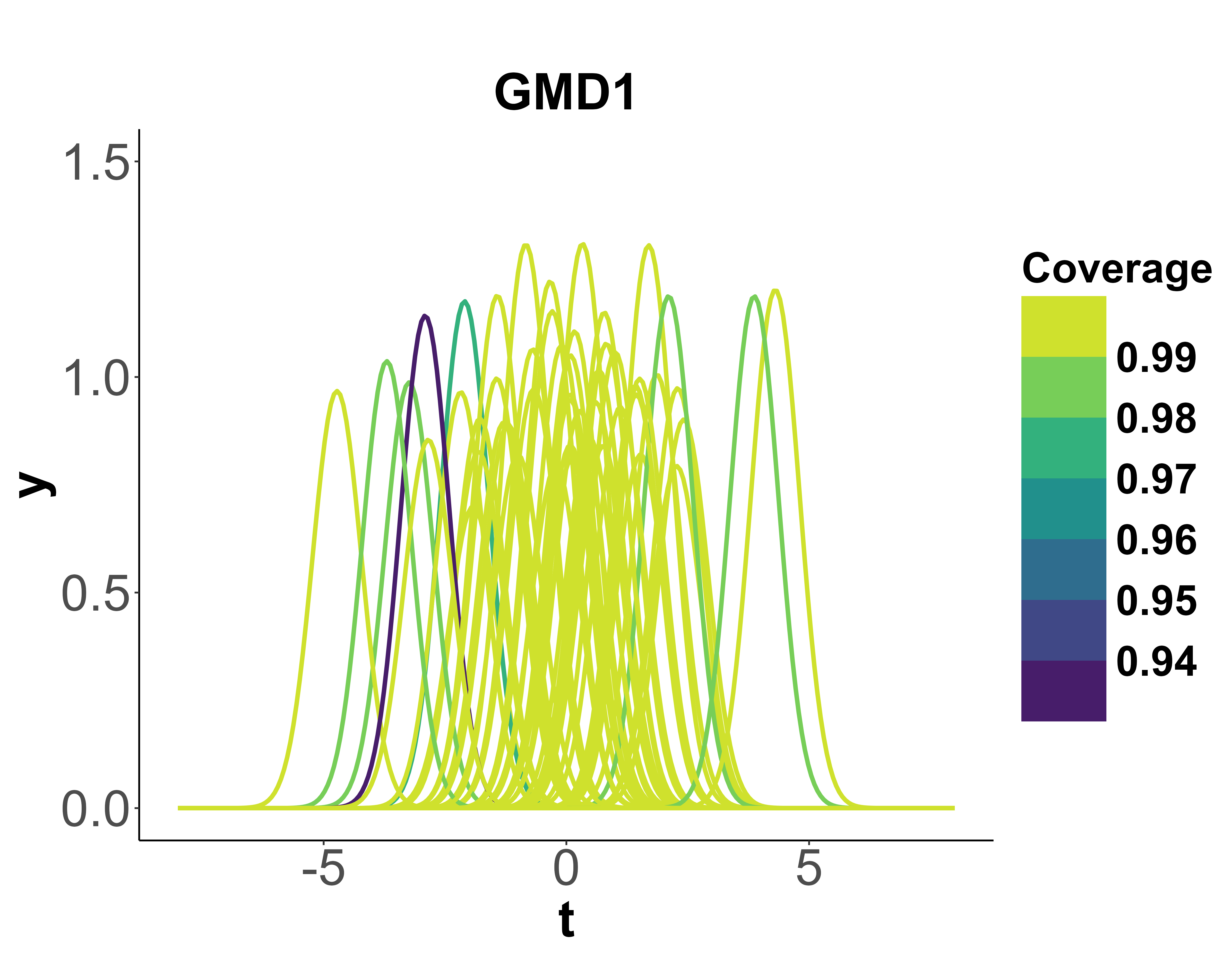} &
      \includegraphics[width=0.50\linewidth]{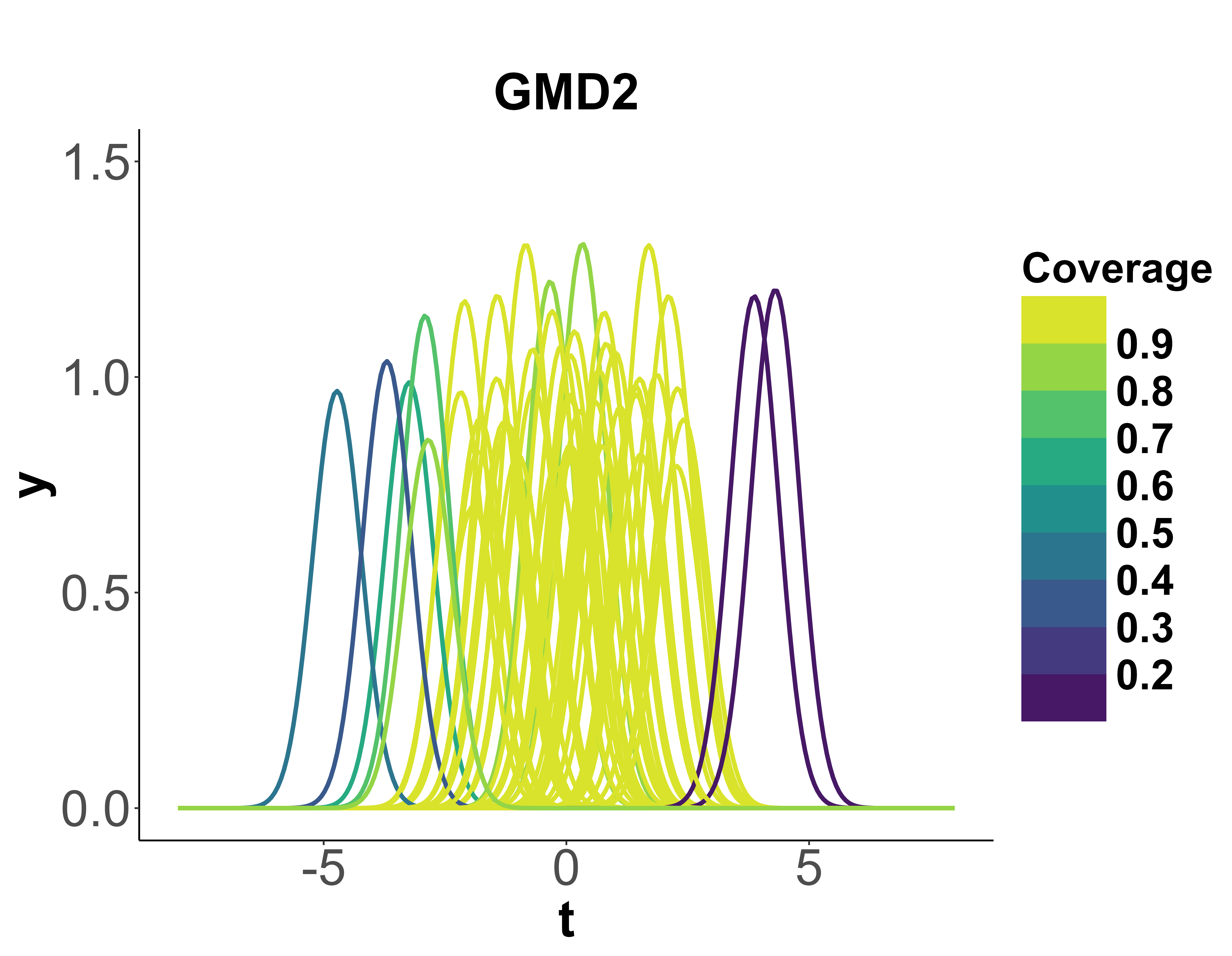} \\
    \end{tabular}
    \caption{Plots of average coverage per test function in the narrow test data for EFDM, SNCM1, GMD1, and GMD2 when $n=100$. Note the radically different color scales between the plots.}
    \label{fig:exp1_covg_nar}
\end{figure}

\begin{figure}[htbp]
    \centering
    \begin{tabular}{cc}
      \includegraphics[width=0.50\linewidth]{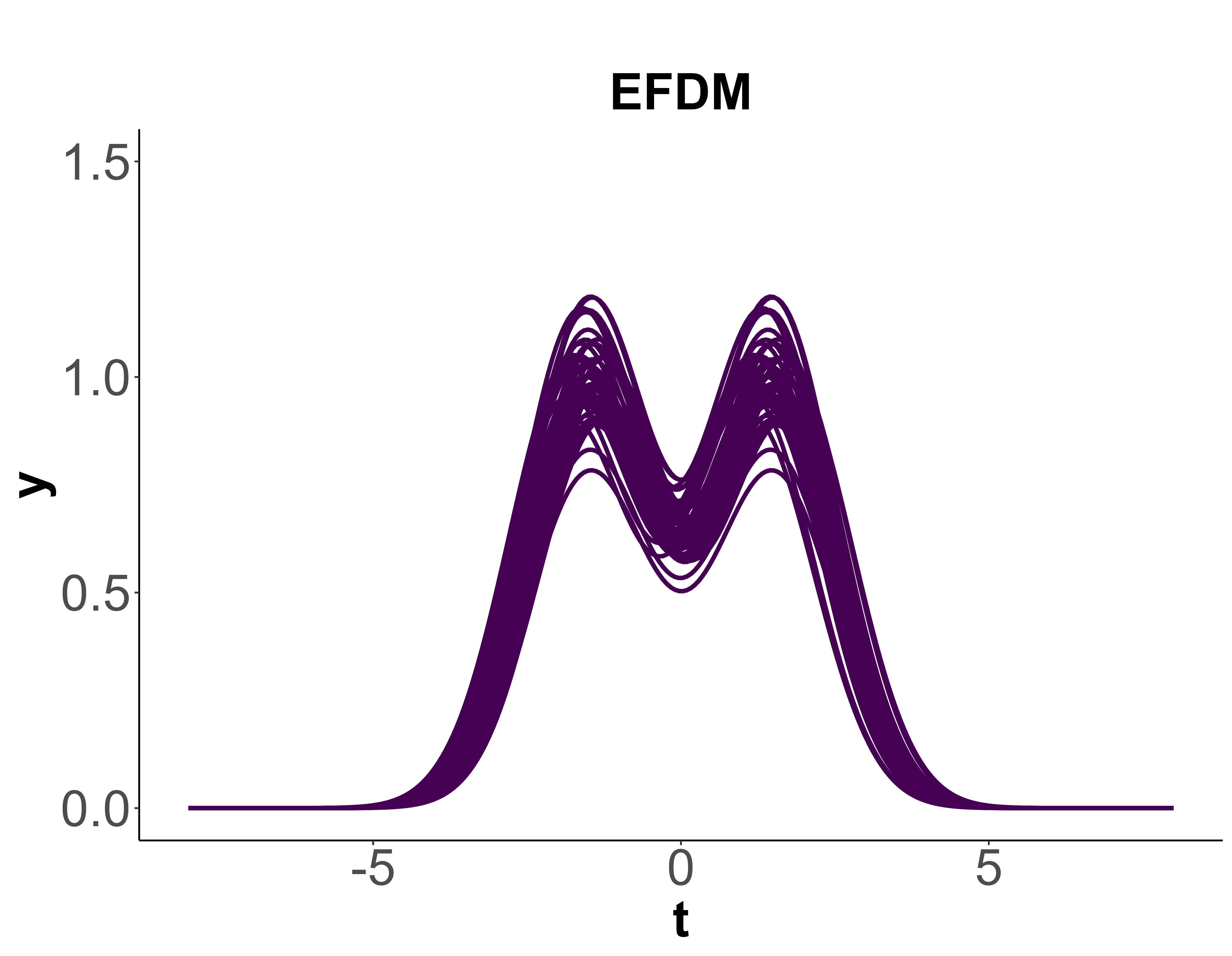} &  
      \includegraphics[width=0.50\linewidth]{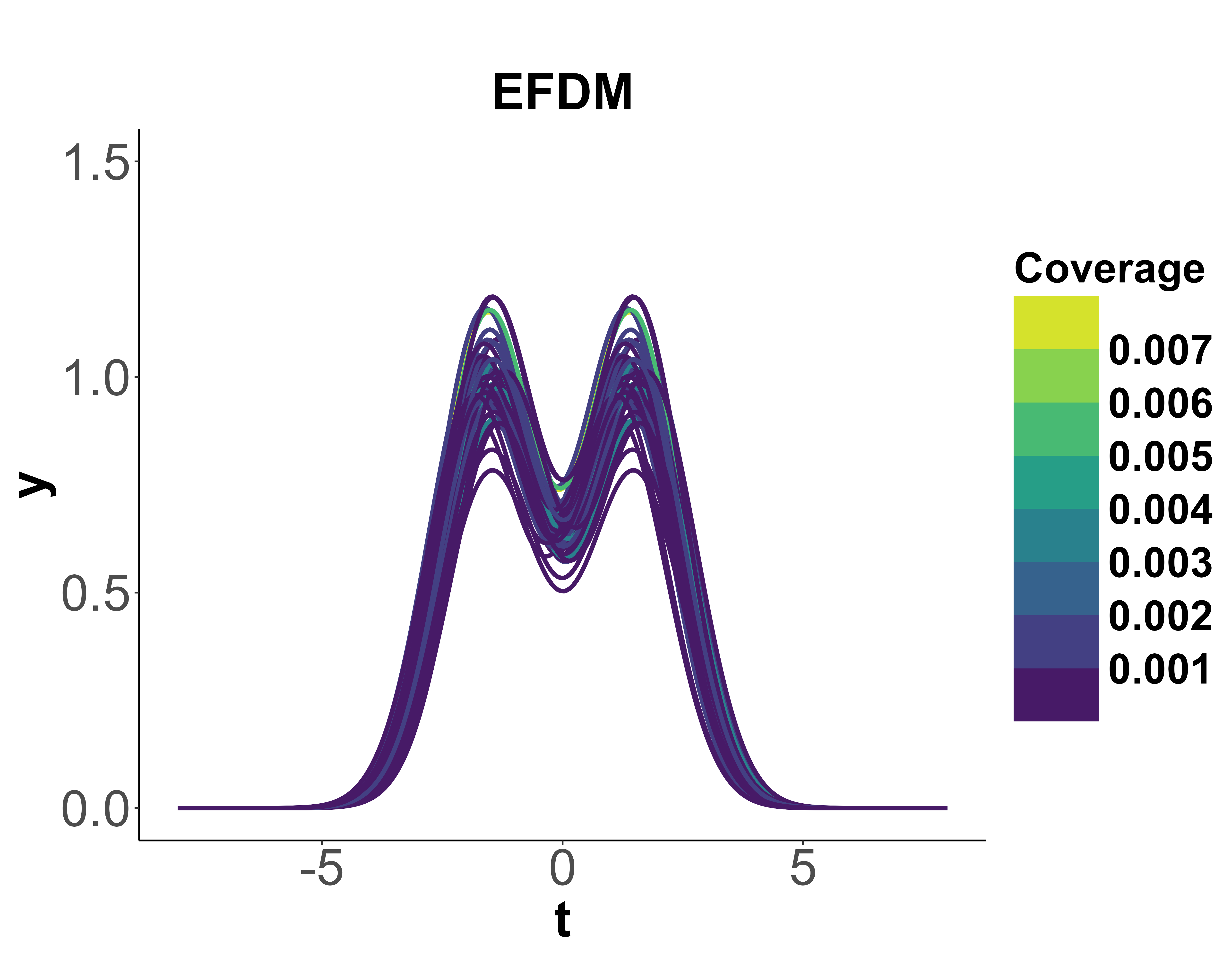} \\
      \includegraphics[width=0.50\linewidth]{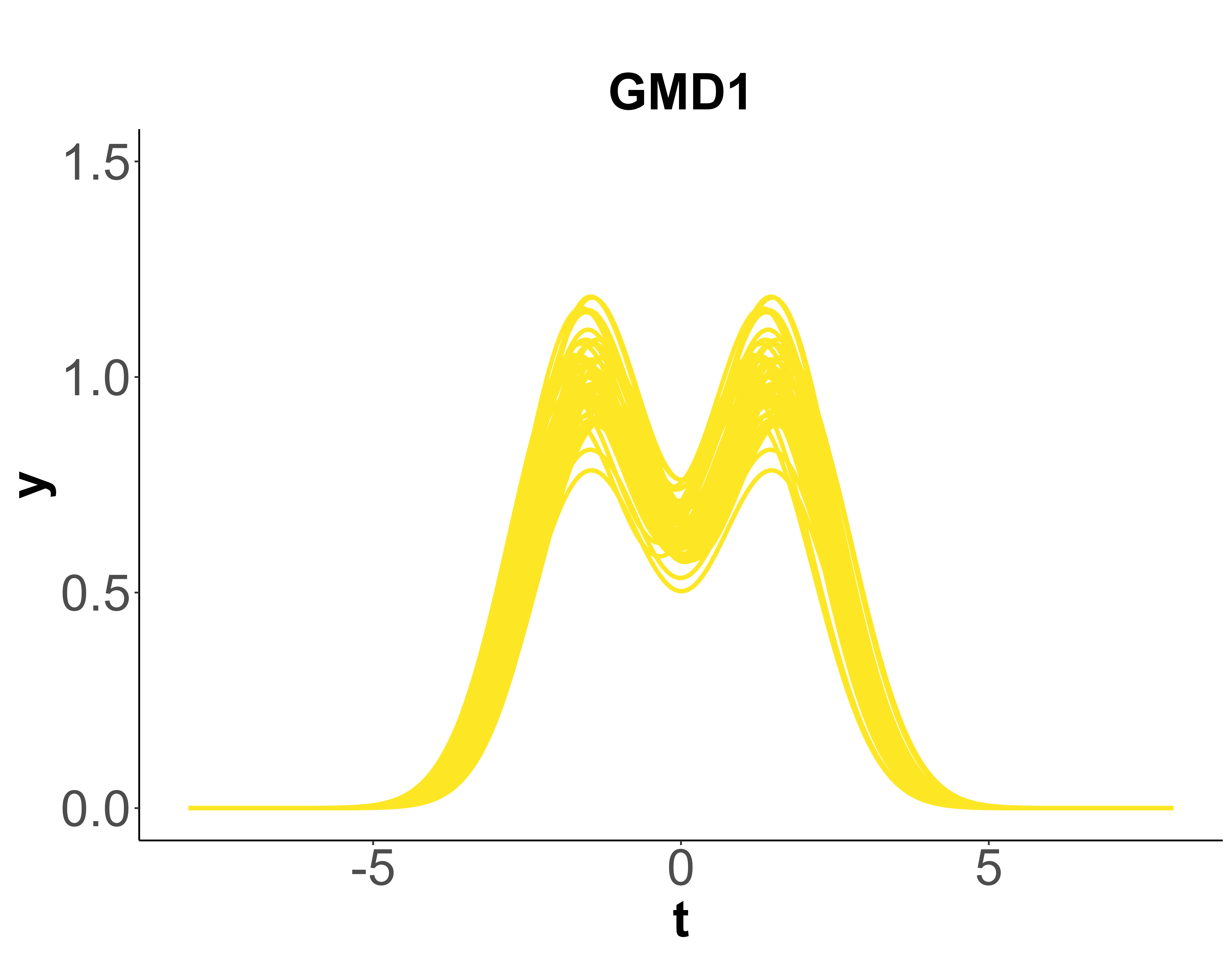} &
      \includegraphics[width=0.50\linewidth]{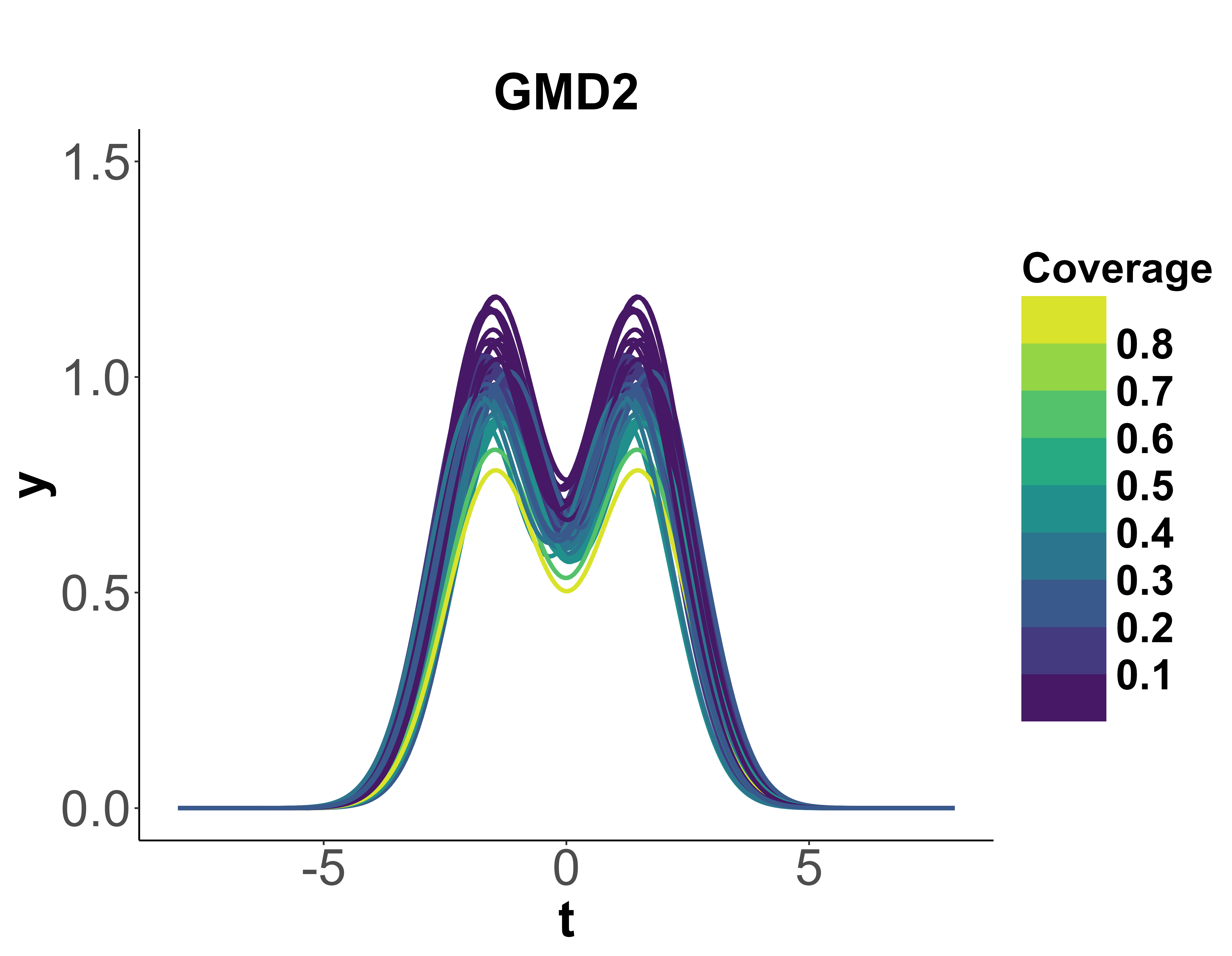} \\
    \end{tabular}
    \caption{Plots of average coverage per test function in the double peaked test data for EFDM, SNCM1, GMD1, and GMD2 when $n=100$. Color scales are not given for EFMD or GMD1 because all coverage values are identical. For EFDM, the coverage values are 0 while for GMD1, they are 1.}
    \label{fig:exp1_covg_dpk}
\end{figure}

Figure \ref{fig:exp1_covg_nar} contains coverage plots, analogous to Figure \ref{fig:exp1_covg}, for the narrow test data set. Similarly, Figure \ref{fig:exp1_covg_dpk} contains coverage plots for the double peaked test data. These plots were produced using EFDM, SNCM1, GMD1, and GMD2 (as described in Table \ref{tab:exp1_methods}) on the 500 full training data sets of size $n = 100$.

\newpage

\subsection{Experiment 2 Metrics}
The tables in this section give the mean and standard deviation values for the true positive rate (TPR), true negative rate (TNR), positive predictive value (PPV), and negative predictive value (NPV) for experiment 2, see Section \ref{sim:exp2} for details of the experiment. Recall that outliers are considered the positive class and inliers the negative class when computing these metrics.

\begin{table}[htbp]
    \centering
    \Large
    \resizebox{\textwidth}{!}{\begin{tabular}{|c|c|c|c|c|c|c|} \hline
      & \multicolumn{3}{|c|}{$5\%$ Outliers in Test Data}   & \multicolumn{3}{|c|}{$10\%$ Outliers in Test Data} \\ \hline
      Method & $n=50$ & $n=100$ & $n=250$ & $n=50$ & $n=100$ & $n=250$ \\ \hline
EFDM & $\textbf{0.854(0.16)}$ & $\textbf{0.997(0.04)}$ & $\textbf{1(0)}$ & $\textbf{0.846(0.12)}$ & $\textbf{0.997(0.03)}$ & $\textbf{1(0)}$ \\ \hline
  SNCM1 & $0.017(0.08)$ & $0.011(0.05)$ & $0.006(0.03)$ & $0.017(0.07)$ & $0.007(0.03)$ & $0.006(0.03)$ \\ \hline
  SNCM2 & $0.024(0.09)$ & $0.015(0.06)$ & $0.006(0.04)$ & $0.022(0.08)$ & $0.013(0.05)$ & $0.007(0.03)$ \\ \hline
  GMD1 & $0.0004(0.009)$ & $0(0)$ & $0(0)$ & $0(0)$ & $0(0)$ & $0(0)$ \\ \hline
  GMD2 & $0.376(0.35)$ & $0.486(0.34)$ & $0.575(0.30)$ & $0.393(0.33)$ & $0.487(0.32)$ & $0.576(0.28)$ \\ \hline
  GMD3 & $-$ & $-$ & $0.554(0.36)$ & $-$ & $-$ & $0.546(0.34)$ \\ \hline
  GMDM & $-$ & $-$ & $0.624(0.34)$ & $-$ & $-$ & $0.629(0.31)$ \\ \hline
    \end{tabular}}
    \caption{Mean(SD) TPR values for experiment 2 using $\alpha = 0.05$.}
    \label{tab:exp2_tpr05}
\end{table}

\begin{table}[htbp]
    \centering
    \Large
    \resizebox{\textwidth}{!}{\begin{tabular}{|c|c|c|c|c|c|c|} \hline
      & \multicolumn{3}{|c|}{$5\%$ Outliers in Test Data}   & \multicolumn{3}{|c|}{$10\%$ Outliers in Test Data} \\ \hline
      Method & $n=50$ & $n=100$ & $n=250$ & $n=50$ & $n=100$ & $n=250$ \\ \hline
  EFDM & $\textbf{0.999(0.01)}$ & $\textbf{1(0)}$ & $\textbf{1(0)}$ & $\textbf{1(0)}$ & $\textbf{1(0)}$ & $\textbf{1(0)}$ \\ \hline
  SNCM1 & $0.042(0.13)$ & $0.032(0.09)$ & $0.022(0.07)$ & $0.04(0.11)$ & $0.024(0.07)$ & $0.02(0.05)$ \\ \hline
  SNCM2 & $0.048(0.13)$ & $0.037(0.1)$ & $0.022(0.07)$ & $0.049(0.11)$ & $0.038(0.09)$ & $0.019(0.05)$ \\ \hline
  GMD1 & $0.0004(0.01)$ & $0(0)$ & $0(0)$ & $0(0)$ & $0(0)$ & $0(0)$ \\ \hline
  GMD2 & $0.629(0.35)$ & $0.781(0.27)$ & $0.866(0.19)$ & $0.63(0.34)$ & $0.782(0.25)$ & $0.86(0.17)$ \\ \hline
  GMD3 & $-$ & $-$ & $0.841(0.24)$ & $-$ & $-$ & $0.839(0.21)$ \\ \hline
  GMDM & $-$ & $-$ & $0.896(0.18)$ & $-$ & $-$ & $0.897(0.15)$ \\ \hline
    \end{tabular}}
    \caption{Mean(SD) TPR values for experiment 2 using $\alpha = 0.10$.}
    \label{tab:exp2_tpr10}
\end{table}

\begin{table}[htbp]
    \centering
    \Large
    \resizebox{\textwidth}{!}{\begin{tabular}{|c|c|c|c|c|c|c|} \hline
      & \multicolumn{3}{|c|}{$5\%$ Outliers in Test Data}   & \multicolumn{3}{|c|}{$10\%$ Outliers in Test Data} \\ \hline
      Method & $n=50$ & $n=100$ & $n=250$ & $n=50$ & $n=100$ & $n=250$ \\ \hline
  EFDM & $0.951(0.05)$ & $0.95(0.04)$ & $\textbf{0.952(0.03)}$ & $0.949(0.05)$ & $0.947(0.04)$ & $0.948(0.03)$ \\ \hline
  SNCM1 & $0.949(0.05)$ & $\textbf{0.951(0.04)}$ & $0.95(0.03)$ & $0.948(0.05)$ & $\textbf{0.95(0.04)}$ & $0.948(0.03)$ \\ \hline
  SNCM2 & $0.95(0.05)$ & $0.95(0.04)$ & $0.949(0.03)$ & $0.949(0.05)$ & $0.948(0.04)$ & $0.947(0.03)$ \\ \hline
  GMD1 & $0.949(0.05)$ & $0.947(0.04)$ & $0.949(0.03)$ & $0.95(0.05)$ & $0.948(0.04)$ & $\textbf{0.949(0.03)}$ \\ \hline
  GMD2 & $\textbf{0.952(0.05)}$ & $0.949(0.04)$ & $0.95(0.03)$ & $\textbf{0.951(0.05)}$ & $0.949(0.04)$ & $\textbf{0.949(0.03)}$ \\ \hline
  GMD3 & $-$ & $-$ & $\textbf{0.952(0.03)}$ & $-$ & $-$ & $\textbf{0.949(0.03)}$ \\ \hline
  GMDM & $-$ & $-$ & $0.944(0.03)$ & $-$ & $-$ & $0.943(0.03)$ \\ \hline
    \end{tabular}}
    \caption{Mean(SD) TNR values for experiment 2 using $\alpha = 0.05$.}
    \label{tab:exp2_tnr05}
\end{table}

\begin{table}[htbp]
    \centering
    \Large
    \resizebox{\textwidth}{!}{\begin{tabular}{|c|c|c|c|c|c|c|} \hline
      & \multicolumn{3}{|c|}{$5\%$ Outliers in Test Data}   & \multicolumn{3}{|c|}{$10\%$ Outliers in Test Data} \\ \hline
      Method & $n=50$ & $n=100$ & $n=250$ & $n=50$ & $n=100$ & $n=250$ \\ \hline
  EFDM & $0.899(0.07)$ & $\textbf{0.899(0.06)}$ & $\textbf{0.901(0.04)}$ & $0.896(0.07)$ & $\textbf{0.897(0.06)}$ & $0.898(0.05)$ \\ \hline
  SNCM1 & $0.900(0.07)$ & $\textbf{0.899(0.06)}$ & $0.900(0.04)$ & $0.899(0.07) $& $\textbf{0.897(0.06)}$ & $0.898(0.04)$ \\ \hline
  SNCM2 & $\textbf{0.901(0.06)}$ & $0.898(0.06)$ & $0.900(0.04)$ & $0.899(0.07)$ & $0.896(0.06)$ & $0.899(0.04)$ \\ \hline
  GMD1 & $0.900(0.08)$ & $0.896(0.06)$ & $0.901(0.05)$ & $\textbf{0.901(0.07)}$ & $\textbf{0.897(0.06)}$ &$ 0.900(0.05)$ \\ \hline
  GMD2 & $0.901(0.07)$ & $0.897(0.06)$ & $\textbf{0.901(0.04)}$ & $\textbf{0.901(0.07)}$ & $0.895(0.06)$ & $\textbf{0.901(0.04)}$ \\ \hline
  GMD3 & $-$ & $-$ & $0.900(0.04)$ & $-$ & $-$ & $0.899(0.05)$ \\ \hline
  GMDM & $-$ & $-$ & $0.881(0.04)$ & $-$ & $-$ & $0.882(0.04)$ \\ \hline
    \end{tabular}}
    \caption{Mean(SD) TNR values for experiment 2 using $\alpha = 0.10$.}
    \label{tab:exp2_tnr10}
\end{table}

\begin{table}[htbp]
    \centering
    \Large
    \resizebox{\textwidth}{!}{\begin{tabular}{|c|c|c|c|c|c|c|} \hline
      & \multicolumn{3}{|c|}{$5\%$ Outliers in Test Data}   & \multicolumn{3}{|c|}{$10\%$ Outliers in Test Data} \\ \hline
      Method & $n=50$ & $n=100$ & $n=250$ & $n=50$ & $n=100$ & $n=250$ \\ \hline
  EFDM & $\textbf{0.588(0.26)}$ & $\textbf{0.584(0.20)}$ & $\textbf{0.567(0.16)}$ & $\textbf{0.708(0.20)}$ & $\textbf{0.716(0.16)}$ & $\textbf{0.708(0.14)}$ \\ \hline
  SNCM1 & $0.007(0.03)$ & $0.007(0.03)$ & $0.003(0.02)$ & $0.013(0.05)$ & $0.009(0.06)$ & $0.008(0.04)$ \\ \hline
  SNCM2 & $0.011(0.04)$ & $0.01(0.04)$ & $0.004(0.02)$ & $0.021(0.08)$ & $0.014(0.05)$ & $0.01(0.04)$ \\ \hline
  GMD1 & $0.0005(0.01)$ & $0(0)$ & $0(0)$ & $0(0)$ & $0(0)$ & $0(0)$ \\ \hline
  GMD2 & $0.282(0.29)$ & $0.331(0.23)$ & $0.385(0.20)$ & $0.426(0.32)$ & $0.5(0.24)$ & $0.558(0.19)$ \\ \hline
  GMD3 & $-$ & $-$ & $0.357(0.23)$ & $-$ & $-$ & $0.496(0.24)$ \\ \hline
  GMDM & $-$ & $-$ & $0.363(0.19)$ & $-$ & $-$ & $0.53(0.19)$ \\ \hline
    \end{tabular}}
    \caption{Mean(SD) PPV values for experiment 2 using $\alpha = 0.05$.}
    \label{tab:exp2_ppv05}
\end{table}

\begin{table}[htbp]
    \centering
    \Large
    \resizebox{\textwidth}{!}{\begin{tabular}{|c|c|c|c|c|c|c|} \hline
      & \multicolumn{3}{|c|}{$5\%$ Outliers in Test Data}   & \multicolumn{3}{|c|}{$10\%$ Outliers in Test Data} \\ \hline
      Method & $n=50$ & $n=100$ & $n=250$ & $n=50$ & $n=100$ & $n=250$ \\ \hline
  EFDM & $\textbf{0.417(0.19)}$ & $\textbf{0.397(0.16)}$ & $\textbf{0.375(0.11)}$ & $\textbf{0.571(0.18)}$ & $\textbf{0.556(0.15)}$ & $\textbf{0.545(0.11)}$ \\ \hline
  SNCM1 & $0.013(0.04)$ & $0.012(0.04)$ & $0.008(0.03)$ & $0.025(0.06)$ & $0.017(0.05)$ & $0.017(0.04)$ \\ \hline
  SNCM2 & $0.014(0.04)$ & $0.015(0.04)$ & $0.009(0.03)$ & $0.031(0.07)$ & $0.027(0.06)$ & $0.016(0.04)$ \\ \hline
  GMD1 & $0.0003(0.01)$ & $0(0)$ & $0(0)$ & 0$(0)$ & $0(0)$ & $0(0)$ \\ \hline
  GMD2 & $0.278(0.20)$ & $0.315(0.15)$ & $0.333(0.10)$ & $0.419(0.23)$ & $0.476(0.14)$ & $0.510(0.11)$ \\ \hline
  GMD3 & $-$ & $-$ & $0.325(0.12)$ & $-$ & $-$ & $0.493(0.12)$ \\ \hline
  GMDM & $-$ & $-$ & $0.297(0.08)$ & $-$ & $-$ & $0.472(0.10)$ \\ \hline
    \end{tabular}}
    \caption{Mean(SD) PPV values for experiment 2 using $\alpha = 0.10$.}
    \label{tab:exp2_ppv10}
\end{table}

\begin{table}[htbp]
    \centering
    \Large
    \resizebox{\textwidth}{!}{\begin{tabular}{|c|c|c|c|c|c|c|} \hline
      & \multicolumn{3}{|c|}{$5\%$ Outliers in Test Data}   & \multicolumn{3}{|c|}{$10\%$ Outliers in Test Data} \\ \hline
      Method & $n=50$ & $n=100$ & $n=250$ & $n=50$ & $n=100$ & $n=250$ \\ \hline
  EFDM & $\textbf{0.992(0.01)}$ & $\textbf{1(0)}$ & $\textbf{1(0)}$ & $\textbf{0.982(0.01)}$ & $\textbf{1(0)}$ & $\textbf{1(0)}$ \\ \hline
  SNCM1 & $0.948(0.004)$ & $0.948(0.003)$ & $0.948(0.002)$ & $0.897(0.01)$ & $0.896(0.004)$ & $0.896(0.004)$ \\ \hline
  SNCM2 & $0.949(0.004)$ & $0.948(0.003)$ & $0.948(0.002)$ & $0.897(0.01)$ & $0.896(0.01)$ & $0.896(0.004)$ \\ \hline
  GMD1 & $0.947(0.003)$ & $0.947(0.003)$ & $0.947(0.002)$ & $0.895(0.01)$ & $0.895(0.005)$ & $0.895(0.003)$ \\ \hline
  GMD2 & $0.967(0.02)$ & $0.973(0.02)$ & $0.977(0.02)$ & $0.936(0.03)$ &$ 0.945(0.03)$ & $0.954(0.03)$ \\ \hline
  GMD3 & $-$ & $-$ & $0.976(0.02)$ & $-$ & $-$ & $0.951(0.04)$ \\ \hline
  GMD4 & $-$ & $-$ & $0.98(0.02)$ & $-$ & $-$ & $0.96(0.03)$ \\ \hline
    \end{tabular}}
    \caption{Mean(SD) NPV values for experiment 2 using $\alpha = 0.05$.}
    \label{tab:exp2_npv05}
\end{table}

\begin{table}[htbp]
    \centering
    \Large
    \resizebox{\textwidth}{!}{\begin{tabular}{|c|c|c|c|c|c|c|} \hline
      & \multicolumn{3}{|c|}{$5\%$ Outliers in Test Data}   & \multicolumn{3}{|c|}{$10\%$ Outliers in Test Data} \\ \hline
      Method & $n=50$ & $n=100$ & $n=250$ & $n=50$ & $n=100$ & $n=250$ \\ \hline
  EFDM & $\textbf{1(0)}$ & $\textbf{1(0)}$ & $\textbf{1(0)}$ & $\textbf{1(0)}$ & $\textbf{1(0)}$ & $\textbf{1(0)}$ \\ \hline
  SNCM1 & $0.947(0.01)$ & $0.946(0.01)$ & $0.946(0)$ & $0.894(0.01)$ & $0.892(0.01)$ & $0.892(0.01)$ \\ \hline
  SNCM2 & $0.947(0.01)$ & $0.946(0.01)$ & $0.946(0)$ & $0.895(0.01)$ & $0.893(0.01)$ & $0.892(0.01)$ \\ \hline
  GMD1 & $0.944(0.01)$ & $0.944(0.004)$ & $0.945(0.003)$ & $0.890(0.01)$ & $0.889(0.01)$ & $0.890(0.01)$ \\ \hline
  GMD2 & $0.98(0.02)$ & $0.988(0.01)$ & $0.993(0.01)$ & $0.959(0.04)$ & $0.975(0.03)$ & $0.984(0.02)$ \\ \hline
  GMD3 & $-$ & $-$ & $0.991(0.01)$ & $-$ & $-$ & $0.981(0.02)$ \\ \hline
  GMDM & $-$ & $-$ & $0.994(0.01)$ & $-$ & $-$ & $0.988(0.02)$ \\ \hline
    \end{tabular}}
    \caption{Mean(SD) NPV values for experiment 2 using $\alpha = 0.10$.}
    \label{tab:exp2_npv10}
\end{table}

\newpage

\subsection{Experiment 3 Metrics}
The tables in this section give the mean and standard deviation TPR, TNR, PPV, and NPV values for experiment 3, see section \ref{sim:exp3}. 

\begin{table}[htbp]
    \centering
    \begin{tabular}{|c|c|c|c|c|} \hline
      &  \multicolumn{2}{|c|}{$\alpha = 0.05$}  & \multicolumn{2}{|c|}{$\alpha = 0.10$} \\ \hline
    Method & $5\%$ Outlier Rate & $10\%$ Outlier Rate & $5\%$ Outlier Rate & $10\%$ Outlier Rate \\ \hline
    EFDM & $\textbf{0.756(0.17)}$ & $\textbf{0.457(0.12)}$ & $\textbf{0.972(0.07)}$ & $\textbf{0.826(0.10)}$ \\ \hline
    SNCM1 & $0(0)$ & $0.002(0.01)$ & $0.014(0.05)$ & $0.010(0.03)$ \\ \hline
    SNCM2 & $0.004(0.03)$ & $0.009(0.03)$ & $0.028(0.08)$ & $0.017(0.04)$ \\ \hline
    GMD1 & $0.078(0.11)$ & $0.007(0.03)$ & $0.224(0.17)$ & $0.050(0.06)$ \\ \hline
    GMDM1 & $0.074(0.11)$ & $0.007(0.03)$ & $0.220(0.16)$ & $0.048(0.07)$ \\ \hline
    GMD2 & $0.252(0.19)$ & $0.064(0.08)$ & $0.510(0.20)$ & $0.141(0.11)$ \\ \hline
    GMDM2 & $0.246(0.18)$ & $0.064(0.07)$ & $0.500(0.21)$ & $0.150(0.10)$ \\ \hline
    GMD3 & $0.150(0.16)$ & $0.022(0.05)$ & $0.308(0.18)$ & $0.049(0.07)$ \\ \hline
    GMDM3 & $0.150(0.15)$ & $0.018(0.04)$ & $0.306(0.20)$ & $0.052(0.08)$ \\ \hline
    \end{tabular}
    \caption{Mean(SD) TPR values for experiment 3.}
    \label{tab:exp3_tpr}
\end{table}

\begin{table}[htbp]
    \centering
    \begin{tabular}{|c|c|c|c|c|} \hline
      &  \multicolumn{2}{|c|}{$\alpha = 0.05$}  & \multicolumn{2}{|c|}{$\alpha = 0.10$} \\ \hline
    Method & $5\%$ Outlier Rate & $10\%$ Outlier Rate & $5\%$ Outlier Rate & $10\%$ Outlier Rate \\ \hline
    EFDM & $\textbf{0.987(0.01)}$ & $\textbf{0.997(0.01)}$ & $\textbf{0.947(0.02)}$ & $\textbf{0.978(0.01)}$ \\ \hline
    SNCM1 & $0.949(0.02)$ & $0.947(0.02)$ & $0.896(0.02)$ & $0.889(0.02)$ \\ \hline
    SNCM2 & $0.949(0.02)$ & $0.944(0.02)$ & $0.897(0.02)$ & $0.888(0.02)$ \\ \hline
    GMD1 & $0.949(0.01)$ & $0.944(0.02)$ & $0.905(0.02)$ & $0.893(0.02)$ \\ \hline
    GMDM1 & $0.950(0.01)$ & $0.945(0.02)$ & $0.904(0.02)$ & $0.895(0.02)$ \\ \hline
    GMD2 & $0.959(0.02)$ & $0.953(0.02)$ & $0.920(0.02)$ & $0.907(0.02)$ \\ \hline
    GMDM2 & $0.959(0.02)$ & $0.952(0.02)$ & $0.917(0.02)$ & $0.905(0.02)$ \\ \hline
    GMD3 & $0.954(0.02)$ & $0.949(0.02)$ & $0.911(0.02)$ & $0.898(0.03)$ \\ \hline
    GMDM3 & $0.954(0.02)$ & $0.949(0.02)$ & $0.910(0.03)$ & $0.899(0.03)$ \\ \hline
    \end{tabular}
    \caption{Mean(SD) TNR values for experiment 3.}
    \label{tab:exp3_tnr}
\end{table}

\begin{table}[htbp]
    \centering
    \begin{tabular}{|c|c|c|c|c|} \hline
      &  \multicolumn{2}{|c|}{$\alpha = 0.05$}  & \multicolumn{2}{|c|}{$\alpha = 0.10$} \\ \hline
    Method & $5\%$ Outlier Rate & $10\%$ Outlier Rate & $5\%$ Outlier Rate & $10\%$ Outlier Rate \\ \hline
    EFDM & $\textbf{0.771(0.15)}$ & $\textbf{0.948(0.10)}$ & $\textbf{0.503(0.08)}$ & $\textbf{0.819(0.10)}$ \\ \hline
    SNCM1 & $0(0)$ & $0.004(0.03)$ & $0.006(0.02)$ & $0.009(0.03)$ \\ \hline
    SNCM2 & $0.003(0.02)$ & $0.016(0.05)$ & $0.013(0.03)$ & $0.015(0.04)$ \\ \hline
    GMD1 & $0.069(0.10)$ & $0.011(0.04)$ & $0.109(0.08)$ & $0.047(0.06)$ \\ \hline
    GMDM1 & $0.068(0.10)$ & $0.012(0.05)$ & $0.107(0.08)$ & $0.046(0.06)$ \\ \hline
    GMD2 & $0.245(0.19)$ & $0.128(0.16)$ & $0.253(0.10)$ & $0.140(0.10)$ \\ \hline
    GMDM2 & $0.240(0.17)$ & $0.129(0.17)$ & $0.244(0.10)$ & $0.149(0.10)$ \\ \hline
    GMD3 & $0.149(0.18)$ & $0.041(0.09)$ & $0.157(0.10)$ & $0.050(0.07)$ \\ \hline
    GMDM3 & $0.150(0.16)$ & $0.035(0.09)$ & $0.153(0.10)$ & $0.053(0.08)$ \\ \hline
    \end{tabular}
    \caption{Mean(SD) PPV values for experiment 3.}
    \label{tab:exp3_ppv}
\end{table}

\begin{table}[htbp]
    \centering
    \begin{tabular}{|c|c|c|c|c|} \hline
      &  \multicolumn{2}{|c|}{$\alpha = 0.05$}  & \multicolumn{2}{|c|}{$\alpha = 0.10$} \\ \hline
    Method & $5\%$ Outlier Rate & $10\%$ Outlier Rate & $5\%$ Outlier Rate & $10\%$ Outlier Rate \\ \hline
    EFDM & $\textbf{0.987(0.01)}$ & $\textbf{0.943(0.01)}$ & $\textbf{0.998(0.004)}$ & $\textbf{0.981(0.01)}$ \\ \hline
    SNCM1 & $0.947(0.001)$ & $0.895(0.002)$ & $0.945(0.003)$ & $0.890(0.004)$ \\ \hline
    SNCM2 & $0.948(0.002)$ & $0.895(0.003)$ & $0.946(0.004)$ & $0.890(0.01)$ \\ \hline
    GMD1 & $0.951(0.01)$ & $0.895(0.003)$ & $0.957(0.01)$ & $0.894(0.01)$ \\ \hline
    GMDM1 & $0.951(0.01)$ & $0.895(0.003)$ & $0.957(0.01)$ & $0.905(0.01)$ \\ \hline
    GMD2 & $0.961(0.01)$ & $0.902(0.01)$ & $0.973(0.01)$ & $0.906(0.01)$ \\ \hline
    GMDM2 & $0.960(0.01)$ & $0.901(0.01)$ & $0.972(0.01)$ & $0.895(0.01)$ \\ \hline
    GMD3 & $0.955(0.0)$ & $0.897(0.01)$ & $0.962(0.01)$ & $0.895(0.01)$ \\ \hline
    GMDM3 & $0.955(0.01)$ & $0.897(0.01)$ & $0.961(0.01)$ & $0.895(0.01)$ \\ \hline
    \end{tabular}
    \caption{Mean(SD) NPV values for experiment 3.}
    \label{tab:exp3_npv}
\end{table}

\end{document}